\documentstyle[sprocl,twoside]{article}  

\def\Journal#1#2#3#4{{#1} {\bf #2}, #3 (#4)}

\def\AP{{\em Ann. Phys.} (N.Y.)}
\def\YF{\em Yad. Fyz.}
\def\NPB{{\em Nucl. Phys.} B}

\def\JHEP{{\em J. High Energy Phys.}}
\def\PLB{{\em Phys. Lett.} B}

\def\PRD{{\em Phys. Rev.} D}
\def\PRP{\em Phys. Rep.}

\def\CMP{\em Comm. Math. Phys.}

\def\IJMPA{{\em Int. J. Mod. Phys.} A}

\def\ra{\rangle}
\def\la{\langle}
\def\dalpha{{\dot{\alpha}}}
\def\dbeta{{\dot{\beta}}}
\def\hph0{{\hphantom{0}}}
\def\c{{\hat{c}}}
\def\r{{\hat{r}}}
\def\K{{\hat{K}}}
\def\J{{\tilde{J}}}
\def\H{{\hat H_\perp}}
\def\overbar{\bar}
\def\blambda{{\overbar{\lambda}}}
\def\bsigma{{\overbar{\sigma}}}
\def\bQ{{\overbar{Q}}}
\def\be{\begin{equation}}
\def\ee{\end{equation}}

\newcommand{\bea}{\begin{eqnarray}}
\newcommand{\eea}{\end{eqnarray}}
\newcommand{\half}{{\scriptstyle{{1\over 2}}}}
\newcommand{\thalf}{{\scriptstyle{{3\over 2}}}}
\newcommand{\fhalf}{{\scriptstyle{{5\over 2}}}}
\newcommand{\shalf}{{\scriptstyle{{7\over 2}}}}

\newcommand{\twoth}{{\scriptstyle{{2\over 3}}}}
\newcommand{\quart}{{\scriptstyle{{1\over 4}}}}
\newcommand{\real}{\relax{\rm I\kern-.18em R}}
\newcommand{\zahlen}{{\rm Z \!\! Z}}

\newcommand{\eff}{{\rm eff}}
\newcommand{\diag}{{\rm diag}}

\newcommand{\veps}{\varepsilon}
\newcommand{\vphi}{\varphi}

\newcommand{\cH}{{\cal H}}
\newcommand{\cO}{{\cal O}}
\newcommand{\cG}{{\cal G}}
\newcommand{\cV}{{\cal V}}
\newcommand{\cJ}{{\cal J}}
\newcommand{\cI}{{\cal I}}
\newcommand{\cL}{{\cal L}}
\newcommand{\cS}{{\cal S}}

\newcommand{\cM}{{\cal M}}
\newcommand{\cY}{{\cal Y}}
\newcommand{\Cite}[1]{$\,$\cite{#1}}
\newcommand{\Ref}[1]{\,(\ref{#1})}

\begin{document}
\sloppy
\hfill{INLO-PUB-05/01}\vskip3mm                                 
\title{THE WITTEN INDEX BEYOND THE ADIABATIC 
APPROXIMATION\footnote{To                                       
appear in the Michael Marinov Memorial Volume,                  
"Multiple facets of quantization and supersymmetry",edited by   
M. Olshanetsky and A. Vainshtein (World Scientific).}           
}
\author{PIERRE VAN BAAL}
\address{Instituut-Lorentz for Theoretical Physics, University of Leiden,\\
P.O. Box 9506, NL-2300 RA Leiden, The Netherlands} 
\maketitle\abstracts{We attempt to deal with the orbifold singularities
in the moduli space of flat connections for supersymmetric gauge theories 
on the torus. At these singularities the energy gap in the transverse 
fluctuations vanishes and the resulting breakdown of the adiabatic 
approximation is resolved by considering the full set of zero-momentum 
fields. These can not be defined globally, due to the problem of Gribov 
copies. For this reason we restrict the fields to the fundamental domain,
containing no gauge copies, but requiring {\em a boundary condition in 
field space}.}
\vspace{0.4cm}          
\vspace{-0.4cm}                                                 
\tableofcontents
\newpage
{\narrower\sl\noindent This paper is dedicated to the memory of \textbf{Michael
Marinov}. He and his family suffered the hardship and humiliation of a 
``refusenik'' in the former Soviet Union. I admired him for his strong moral 
principles and persistence. I feel fortunate having known him, and having 
experienced his kindness. Michael has done pioneering work involving Grassmann 
variables, supersymmetry, geometric quantization and quantum tunneling. I hope 
the following result would have been to his liking. My attempts to address this 
problem stem from the period I first met Michael at a workshop in Trieste, 
now just over 10 years ago.\par}
\section{Introduction}\label{sec:intro}
We revisit supersymmetric Yang-Mills theories on the torus to study the vacuum 
state in connection with the Witten index.\cite{WiIn} The torus geometry is 
crucial to preserve the supersymmetry. The index counts the number of 
quantum states (fermionic states with a negative sign). Due to supersymmetry, 
states at non-zero energy occur in fermionic and bosonic pairs, and do not 
contribute to the Witten index. The counting can therefore be reduced to the 
vacuum sector. The Hamiltonian is given by $H=\half\{Q,Q^\dagger\}$ with $Q$, 
$Q^\dagger$ the supersymmetry generators, and unbroken supersymmetry requires 
these to annihilate the vacuum state, hence giving a zero vacuum energy. Only
at zero energy there can be an absence of full pairing, which would make the 
Witten index non-zero. A zero value of this index thus indicates supersymmetry
may be spontaneously broken.

In perturbation theory bosonic (gluon) loops are canceled by fermionic (gluino)
loops, and applied to the problem of non-abelian gauge theories in a finite 
volume, it leads to the absence of an induced effective potential on the 
moduli space of flat connections (the so-called vacuum valley). 
The gluinos are represented as Weyl fermions in the adjoint representation of 
the gauge group, denoted by $\lambda_\alpha^a$, with $\alpha$ a two-component 
spinor index. They are the superpartners of the gluons. 

A technical problem that has remained unresolved ever since Witten's original 
work is associated to a breakdown of the adiabatic approximation in the 
reduction of the degrees of freedom to those of the classical vacuum, when 
using periodic boundary conditions. This classical vacuum is defined up to a 
gauge transformation by the set of zero-momentum abelian gauge fields. Its 
gauge invariant parametrization is in terms of the Wilson loops that wind 
around the three compact directions of the torus, which are {\em compact}
variables. This describes the vacuum valley as an orbifold, $T^3/Z_2$ for
SU(2). The orbifold singularities arise where the flat connection is 
invariant under (part of) the Weyl group (the remnant gauge transformations 
that leave the set of zero-momentum abelian gauge fields invariant). For 
SU(2) their are eight orbifold singularities (related to $A=0$ by {\em 
anti-periodic} gauge transformations).

Without the contributions from the fermions, the wave function would be 
localized to these orbifold singularities. The singularity is resolved by 
including all the zero-momentum gauge fields.\cite{Bjor,Lue1} A reduction near 
$A=0$ in terms of the abelian modes is impossible, since here the energy gap 
between the fluctuations in the abelian and the non-abelian field directions 
vanishes. This is the source of {\em singular} non-adiabatic behavior, and 
remains so for the supersymmetric case.\cite{PvBG} Resolving the orbifold 
singularity in the supersymmetric case leads one to consider the reduction 
to zero (spatial) dimensions.\cite{ClHa,Smil} In the context of the 
supermembrane\Cite{Mem} it was found that the spectrum is continuous, down to 
zero-energy. One can construct trial wave functions with arbitrarily small 
energy,\cite{LNdW} by moving its support far from $A=0$. In the case of the 
supermembrane the vacuum valley is {\em non-compact}. For gauge theories on 
the torus, the compactness of the vacuum valley arises due to identifications 
under periodic gauge transformations with non-zero momentum. Such a gauge 
transformation does {\em not preserve the momentum} for the non-abelian modes. 
What is zero-momentum near $A=0$, is non-zero momentum near a gauge copy of 
$A=0$. We therefore have to match, somehow, the behavior {\em near each} of 
the orbifold singularities to the behavior {\em far away}, where a reduction 
to the abelian zero-momentum modes is dynamically justified, and where the 
Hamiltonian is just the standard Laplacian on the torus. 

\subsection{Status quo}\label{subsec:quo}
Let us first summarize the current state of affairs. Assuming the reduction 
to the vacuum valley is justified, we note that the zero-momentum gluinos 
associated with the abelian generators, each with two helicities, carry no 
energy, which is the source of the vacuum degeneracy.\cite{WiIn} These gluinos 
have to be combined in Weyl invariant combinations, respecting Fermi-Dirac 
statistics. There are $r$ (with $r$ the rank of the gauge group) independent 
invariants, made from
\be
U=\delta_{ab}\epsilon^{\dalpha\dbeta}\blambda_\dalpha^a\blambda_\dbeta^b
\ee
and its powers. So one has $U^n|0\ra$, $n=0,1,\cdots,r$, as bosonic vacuum 
states, and no invariant fermionic vacuum states. This led Witten to an index 
equal to the rank of gauge group plus one, $r+1$. To circumvent the problems 
near the orbifold singularities, Witten\Cite{WiIn} considered the alternative 
of twisted boundary conditions.\cite{Tho1} For SU($N$) the same result for the 
index follows. However, other groups do in general not admit the type of 
twisted boundary conditions that completely remove the continuous vacuum 
degeneracy (with its associated orbifold singularities). In particular these 
twisted boundary conditions could not be used to resolve a discrepancy with 
the infinite volume result based on the determination of the gluino condensate 
through instanton contributions,\cite{NSVZ,ShVa,AmVe} relying on the fact 
that the index should not change with the volume, or for that matter any 
other smooth deformation. 

The gluino condensate calculation has been justified by first adding matter 
fields, which introduce an external mass scale so as to control the weak 
coupling expansion. One then relies on the index being constant under a 
smooth deformation (through holomorphy), that decouples the extra matter 
sector.\cite{ShVa} In the direct (strong coupling) approach, since the 
instantons have more than two gluino zero modes, 
it seems the condensate $\la\lambda\lambda\ra$ vanishes. Instead the 
appropriate power of the gluino condensate, $\la(\lambda\lambda)^h\ra$, is 
considered where $h$ is the so-called dual Coxeter number of the gauge group, 
$h=N$ for SU($N$), which counts the number ($2h$) of gluino zero modes. 
Invoking the cluster decomposition property, it is this power in $h$ that 
gives the number of vacuum states, 
\be
\la\lambda\lambda\ra\equiv e^{2\pi in/h}\left(|\la(\lambda\lambda
)^h\ra|\right)^{1/h},\quad n=1,2,\cdots,h. 
\ee
These arguments seem reasonable, but are not rigorous,\cite{AmVe,Shif} and
suffer from a discrepancy in the prefactor ($\sqrt{5/4}$ for SU(2)). More
recently, use has been made of the constituent nature of periodic instantons
(or calorons),\cite{KrvB,LLYi} in the context of a Kaluza-Klein reduction
with periodic gluinos (as opposed to a high temperature reduction with
anti-periodic gluinos, which would break the supersymmetry). The constituent 
monopoles, with $A_0$ playing the role of a Higgs field, have exactly two 
zero-modes and saturate the condensate, $\la\lambda\lambda\ra$, giving the
correct prefactor.\cite{Khoz} In this case it is the compactification scale 
that justifies the semiclassical approximation.

The mismatch in the Witten index between small and infinite volumes occurs 
for SO($N>6$) and the exceptional groups. There has, however, been a recent 
revision in counting the number of vacuum states in a finite volume. In a study
of D-brane orientifolds in string theory, Witten\Cite{WiBr} constructed for 
SO(7) an extra disconnected component on the moduli space of flat gauge 
connections, which can be embedded easily in SO($N>7$). For SO(7) and SO(8) 
this gives an isolated component of the moduli space, contributing only one 
extra vacuum state. For SO($N>8$) the extra component in the moduli space
behaves like the trivial component for SO($N$-7). Adding $r+1$ coming from the 
SO($N$) and SO($N-7$) moduli space components gives the dual Coxeter number of 
SO($N$), thereby yielding the same number of vacuum states as obtained in the 
infinite volume. 

Witten's construction based on orientifolds does not work for the exceptional 
groups. This naturally led to a derivation of the extra vacuum states in a 
field theoretic context,\cite{KeRS} trivially extended to the exceptional 
group $G_2$, as a subgroup of SO(7). It was subsequently solved for other 
exceptional groups with periodic boundary conditions\Cite{Keur,KaSm} and 
for any group with twisted boundary conditions.\cite{BoFM} Twisted boundary 
conditions usually do not remove all the vacuum degeneracies, but it is 
important that the number of vacuum states is independent of the twist for 
all gauge groups that have a non-trivial center. The origin of the extra 
moduli space components is actually not too hard to understand.\cite{Keur} 
Large gauge groups can have subgroups that are products of unitary groups, 
which each would allow for twisted boundary conditions. By choosing ``twists" 
from all subgroups to cancel (or give the desired total ``twist"), one obtains 
flat connections that can not be deformed to the Cartan subalgebra (which 
supports the trivial component of flat connections). 

Although these new results for counting the number of vacuum states in a 
finite volume remove the urgency of addressing the problem with the adiabatic 
approximation, it does remain a sore point in the finite volume analysis, as 
stressed again by Witten.\cite{WiRe} On the one hand the wave function
on the vacuum valley has to be constant, on the other hand it seems to 
want to vanish near the orbifold singularities. We will argue that 
deviations from a constant will be confined to a distance from each 
orbifold singularity that is $\cO(g^{2/3}(L))$ times the distance between 
the orbifold singularities, where the dependence on $L$ is due to the 
asymptotically free non-trivial running of the coupling, appropriate for 
$N=1$ supersymmetric gauge theories. At the same time, however, in this
small region the energy needs to remain zero and here lies the burden of 
the proof.

\subsection{The adiabatic approximation}\label{subsec:appr}
\begin{figure}[htb]
\vspace{5.6cm}
\includegraphics{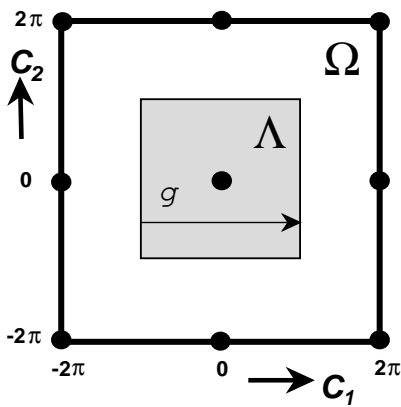}
\caption{A two dimensional slice of the vacuum valley along the $(C_1,C_2)$
plane. The grey square is the fundamental domain $\Lambda$. The dots are 
gauge copies of the origin (which turn out to lie on the Gribov horizon 
$\Omega$, indicated by the fat square).}\label{fig:torus}
\end{figure}

We are interested in constructing the vacuum wave function in sufficiently 
small volumes. Our convention is to choose the dependence on the bare coupling
constant such that it appears as an overall factor $1/g^2_0$,
\be
\cL=-\frac{1}{4g_0^2}(F^a_{\mu\nu})^2+\frac{i}{2g_0^2}
     \blambda^a\gamma_\mu(D^\mu\lambda)^a.
\ee
The reduction to the zero-momentum degrees of freedom, as in the bosonic case, 
will replace the bare coupling constant by a running and asymptotically free 
coupling constant $g(L)$. The zero-momentum gauge fields are parametrized as 
$A_i=ic_i^a\tau_a/(2L)$, with $\vec\tau$ the Pauli matrices. 
The vacuum valley is parametrized by the abelian degrees of freedom. These are 
defined by $r_i$, with $r_ir_j=\sum_a c_i^ac_j^a$, for each $i$ and $j$. 
Alternatively, we may parametrize the vacuum valley by $r_i=C_i\equiv c_i^3$, 
choosing the maximal abelian subgroup to be generated by the third Pauli 
matrix $\tau_3$. The effect of the periodic gauge transformations,
\be
g_{\vec n}(\vec x)=\exp(-2\pi i\vec n\cdot\vec x\tau_3/L),
\ee 
is to shift $\vec C$ over $4\pi\vec n$, making the vacuum valley into a 
torus, see Fig.~\ref{fig:torus}.

Also in the supersymmetric case, the Hamiltonian is invariant under
anti-periodic gauge transformations (gauge transformations periodic up 
to an element of the center of the gauge group). When $\vec n$ has at
least one of its components half integer it is homotopically non-trivial.
Since it is a symmetry of the full Hamiltonian, this is one way to see
that the vacuum valley has 8 orbifold singularities, related to $A=0$ by 
shifts over {\em half-periods} ($2\pi$ in each of the three directions). 
Alternatively, the action of the Weyl group on the vacuum valley is given 
by the reflection $\vec C\rightarrow -\vec C$, and the orbifold singularities 
are at the eight fixed points of this symmetry (combined with the shift 
symmetries). The cell $\vec C\in[-\pi,\pi]^3$ can be used as a fundamental 
domain $\Lambda$, see Fig.~\ref{fig:torus}. Any point on the vacuum valley 
can be reached by applying suitable gauge transformations. Opposite sides 
on it's boundary are identified under these homotopically non-trivial
gauge transformations. The representations of their homotopy define the 
electric flux quantum numbers as introduced by 't Hooft.\cite{Tho1} We will 
here only consider the sector with zero electric flux, i.e. the trivial 
representation, where wave functions at opposite sides are equal. 

We need to reconsider the construction of the effective Hamiltonian, since the 
gluino loops tend to cancel the gauge loops. For a background with zero field
strength, the effective potential is easily seen to vanish to all orders in 
perturbation theory.\cite{WiIn} But to resolve the orbifold singularity near
$A=0$, we do not wish to integrate out the non-abelian zero-momentum modes.
These modes can have non-zero field strength, and the quantum corrections for 
these are expected not all to cancel. Otherwise the $\beta$-function would 
vanish, which we know not to be the case for $N=1$ supersymmetric gauge 
theories. If necessary, field redefinitions should restore the supersymmetry. 

In the background field gauge the one-loop effective potential reads
\be
V_1(c)=L^{-1}\left\{\frac{1}{4}\left(\frac{L^{d-3}}{g_0^2}+\alpha_2(d)\right)
(F^a_{ij})^2+\alpha_3 (F^a_{ij})^2(c_k^b)^2+\alpha_5{\det}^2(c)+\cdots\right\},
\ee
where $F^a_{ij}\equiv-\veps_{abd}c_j^bc_k^d$. The coefficients (labeled as 
in the bosonic case\Cite{Vba4}) in dimensional regularization are, up to terms 
vanishing at $d=3$, given by ($\vec k\in(2\pi\zahlen)^3$) 
\be
\alpha_2(d)=-\frac{(d+3)(d+6)}{12d}\sum_{\vec k\neq\vec 0}\frac{1}{|\vec k|^3},
\ \alpha_3=-\frac{1}{32}\sum_{\vec k\neq\vec 0}\frac{1}{|\vec k|^5},
\ \alpha_5=-\frac{15}{16}\sum_{\vec k\neq\vec 0}\frac{1}{|\vec k|^5}.
\ee
The result for dimensional reduction (vector and spinor indices strictly in 4 
space-time dimensions) is simply obtained here (and below) by putting $d=3$, 
or\Cite{Lue1} $\alpha_2(3)=-\frac{3}{2}\sum_{\vec k\neq\vec 0}|\vec k|^{-3}=
\frac{3}{4\pi^2(d-3)}+\frac{3}{8\pi^2}\left(\frac{1}{11}+0.409052802\cdots
\right)+\cO(d-3)$. The effective potential vanishes along the vacuum valley, 
as expected. 

In minimal subtraction one defines 
\be
\frac{L^{d-3}}{g_0^2}=-\frac{3}{4\pi^2(d-3)}+\frac{1}{g^2(L)},\quad 
\frac{1}{g^2(L)}\equiv-\frac{3}{4\pi^2}\ln(L\Lambda_{{}_{\rm MS}}), 
\ee
such that $L^{d-3}/g_0^2+\alpha_2(d)=1/g^2(L)+\alpha_2$, with $\alpha_2$ the 
finite part of $\alpha_2(d)$, and $g(L)$ the running coupling appropriate
for the supersymmetric theory. The ``electric" part of the effective
Lagrangian is also not difficult to compute in the background field 
calculation,
\be 
\half L\left(\frac{L^{d-3}}{g_0^2}+\alpha_1(d)\right)\left((D_0 c_i)^a(t)
\right)^2,\quad\alpha_1(d)=-\frac{3+5d}{4d}\sum_{\vec k\neq\vec 0}
\frac{1}{|\vec k|^3}.
\ee
With $(D_0c_i)^a=\dot c_i^a+\veps_{adb}c_0^dc_i^b$, we keep $c^a_0\neq0$
to preserve supersymmetry,
\be
\delta A_\mu^a=\frac{i}{2}(\overbar\veps\gamma_\mu\lambda^a-\blambda^a
\gamma_\mu\veps),\quad\delta\lambda^a=\gamma^{\mu\nu}F_{\mu\nu}^a\veps,\quad
\delta\blambda^a=-\overbar\epsilon\gamma^{\mu\nu}F_{\mu\nu}^a,
\ee
which allows for a consistent truncation to the zero-momentum sector. 
Both in dimensional regularization and dimensional reduction the finite parts,
$\alpha_{1,2}$, are equal and will be absorbed in the running coupling. 
Finally, for the gluino part of the effective Lagrangian we find
\be
\frac{iZ}{2g_0^2}L^{d-1}\blambda^a(t)\left(L\gamma_0\dot\lambda^a(t)+\kappa
\veps_{adb}c_i^d(t)\lambda^b(t)\right),\ Z=1+\frac{g_0^2}{2}
\sum_{\vec k\neq\vec 0}\frac{1}{|\vec k|^3},\ \kappa=1.
\ee
In the background field gauge only the gluino field requires a rescaling
to restore the supersymmetry (with our use of component fields, implying
the Wess-Zumino gauge, we do not maintain explicit supersymmetry, but the 
final results should of course not depend on this). 

We also used the Lorentz gauge, which shifts the coefficients by\Cite{Vba4} 
$\delta\{\alpha_1,\alpha_2\}=\{1,2\}\sum_{\vec k\neq\vec 0}|\vec k|^{-3}$, 
$\delta\{\alpha_3,\delta\alpha_5\}=\{\frac{7}{192},-\frac{57}{32}\}\sum_{\vec 
k\neq\vec 0}|\vec k|^{-5}$. In addition $\delta Z_2=0$, and $\delta\kappa=\half
\alpha_1$. Infinities are absorbed by a simple rescaling of the gluon field,
as guaranteed by gauge invariance. The changes in $\alpha_3$ and $\alpha_5$ can
be absorbed by a finite but non-linear field redefinition.\cite{Vba4} All we 
want to stress here is, like in the abelian case,\cite{ASmi} that supersymmetry does not imply that the corrections (like $\alpha_3$ and 
$\alpha_5$) need to vanish, but a self-consistent calculation requires further 
terms in the effective action, involving a non-static background field (for 
the bosonic case most reliably done in a Hamiltonian setting\Cite{Vba4}). 
The importance of these
higher order terms will become clear later. To lowest order, one finds the 
Hamiltonian truncated to the zero-momentum sector, with the bare coupling 
replaced by the renormalized coupling. 

The energy gap in the fluctuations transverse to the vacuum valley is easily
read off from the Lagrangian. Close to the origin it is given by $2|\vec C|/L$.
Integrating out transverse degrees of freedom is only reliable if the energy of
the low-lying states is smaller than this gap. This energy behaves as 
$g^{2/3}(L)/L$ (as shown in the next section, rescaling $c$ with $g^{2/3}(L)$ 
makes the Hamiltonian proportional to $g^{2/3}(L)/L$). The gap in the 
transverse fluctuations thus becomes of this order where $|\vec C|$ is 
of order $g^{2/3}(L)$. 

Consider a sphere of radius $g^{1/3}(L)$ around each orbifold singularities,
beyond which the adiabatic approximation is accurate. There, the wave function 
can be reduced to the vacuum valley and, assuming indeed it has zero energy, 
it will be constant. (Note that the chosen parametrization of the vacuum valley 
is independent of $g(L)$ and $L$.) As long as $g(L)$ is small, we may assume 
the wave function in the neighborhood of the orbifold singularity to respect 
spherical symmetry. In the bosonic case, once the wave function spreads out
over the vacuum valley, any spherical symmetry is quickly lost. In the 
supersymmetric case, however, the reduced wave function on the vacuum valley
will become constant up to exponential corrections at distances much greater 
than $g^{2/3}(L)$ from the orbifold singularities, which are separated over a 
distance $2\pi$. (Rescaling $c$ with $g^{2/3}(L)$, the boundary of the sphere 
introduced above is at $g^{-1/3}(L)$, with the boundary of the fundamental 
domain and the other orbifold singularities an other factor $g^{-1/3}(L)$ 
removed from that.) Instead of insisting that the groundstate wave function 
is normalizable, we should rather insist on its projection to the vacuum valley
to become constant. As long as the wave function is bounded everywhere, the 
issues of {\em normalizability}\Cite{RecM} in the context of the supermembrane 
or M-theory\Cite{BFSS} applications, is therefore of no relevance here. But the
challenge of explicitly constructing the ground state, of course, remains the 
same (a subtle difference will be pointed out later).

As we have argued, when the volume is small enough, we may assume the
groundstate wave function of the effective Hamiltonian to be spherically
symmetric. The boundary conditions at the boundary of the fundamental domain 
are replaced by requiring this wave function to approach a constant, after 
projecting to the vacuum valley. This projection is well defined, because 
at large separations from the origin, the wave function becomes exponentially 
localized transverse to the vacuum valley. Spherical symmetry near the 
orbifold singularities will dramatically simplify the analysis. 

In the next section we will set up the zero-momentum supersymmetric 
Hamiltonian, and its reduction to the gauge invariant, spherically symmetric 
sector. This is not new,\cite{Itoy,Halp} but we will be able to push it to 
the point where we can explicitly construct a complete basis of states that 
respect these symmetries. The Hamiltonian can be split into a ``radial" and 
``angular" part, and our basis explicitly diagonalizes the angular part 
(``spherical harmonics") in terms of invariant polynomials. This may be 
useful in a more general context.

\section{The Hamiltonian}\label{sec:Ham}

The conventions for the superalgebra we follow are those of Wess and 
Bagger.\cite{Wess} For the formulation of the supersymmetric Hamiltonian 
we follow to a large extent earlier work.\cite{Itoy,Halp,Froh} We start from 
the supercharge operators, 
\be
Q_\alpha=\sigma^j_{\alpha\dbeta}\blambda_a^\dbeta\left(-i\frac{\partial}{
\partial V^j_a}-iB_j^a\right),\quad\bQ_\dalpha=\lambda_a^\beta\sigma^j_{
\beta\dalpha}\left(-i\frac{\partial}{\partial V_j^a}+iB_j^a\right),
\ee
with $V_i^a\equiv c_i^a/(g(L)L)$ and $\sigma^j=\tau^j$ (and $\sigma^0$ the 
unit) as $2\times2$ matrices. Restricting to the zero-momentum modes, both the 
Weyl spinors $\lambda_a^\beta$ and $\blambda_a^\dbeta$ are constant. Lowering 
indices is done with $\epsilon_{\alpha\beta}=\epsilon_{\dalpha\dbeta}=-i
\tau_2$, $\delta_{ab}$ and $\eta_{\mu\nu}=\diag(1,1,1,-1)$ (or $\delta_{ij}$) 
for respectively the spinor, group and space-time (or space) indices, and 
raising of indices is done with the inverse of these matrices. Repeated
indices are assumed to be summed over, but to keep notations transparent 
we will not always balance the positions of the gauge and space indices. 
For zero-momentum gauge fields
\be
B_i^a=-\half g\veps_{ijk}\veps_{abc}V_j^bV_k^c.
\ee

In the Hamiltonian formulation the anti-commutation relations
\be
\{\lambda^{a\alpha},\blambda^{b\dbeta}\}=\bsigma_0^{\dbeta\alpha}\delta^{ab},
\quad\{\lambda^{a\alpha},\lambda^{b\beta}\}=0,\quad
\{\blambda^{a\dalpha},\blambda^{b\dbeta}\}=0,
\ee
with $\bsigma_0$ the unit $2\times2$ matrix (one has $(\bsigma^\mu)^{\dalpha
\alpha}=\epsilon^{\dalpha\dbeta}\epsilon^{\alpha\beta}(\sigma^\mu)_{\beta\dbeta}
$), give
\be
\{Q_\alpha,\bQ_\dalpha\}=2(\sigma_0)_{\alpha\dalpha}\cH-
2(\sigma^i)_{\alpha\dalpha}V_i^a\cG_a,
\ee
where 
\be
\cG_a=ig\veps_{abc}\left(V_j^c\frac{\partial}{\partial V_j^b}-
\blambda^b\bsigma_0\lambda^c\right)
\ee
is the generator of infinitesimal gauge transformations, and 
$\cH$ is the Hamiltonian density
\be
\cH=-\half\frac{\partial^2}{\partial V_i^a\partial V_i^a}+
\half B_i^aB_i^a-ig\veps_{abc}\blambda^a\bsigma^j\lambda^b V_j^c.
\ee
Splitting the Hamiltonian, $\int d^3x\cH\equiv g^{2/3}(L)H/L$, in its bosonic 
and fermionic pieces, $H=H_B+H_f$, we find with $c_i^a=g^{2/3}(L)\c_i^a$
\be
H_B=-\half\left(\frac{\partial}{\partial \c_i^a}\right)^2+\half\left(\hat B_i^a
\right)^2,\quad H_f=-i\veps_{abd}\blambda^a\bsigma^i\lambda^b\c_i^d,
\ee
where
\be
\hat B^i_a=-\half\veps^{ijk}\veps_{abd}\hat c_j^b\c_k^d.
\ee

As discussed in the previous section, the orbifold singularities, other than 
at $\c=0$, lie at a distance $2\pi g^{-2/3}(L)$ in these new variables $\c$ 
(measured along the vacuum valley where $\hat B$ vanishes). We want to solve 
for the groundstate wave function such that for $|\c|\gg 1$ it becomes a 
constant, after projecting to the vacuum valley (we will come back to this 
projection later). As this boundary condition is compatible with spherical 
symmetry, i.e. it goes to the same constant for all directions on the vacuum 
valley, we will restrict ourselves to wave functions $\Psi(\c)$ that are 
spherically symmetric and gauge invariant. We stress this is an accidental
spherical symmetry, that holds in sufficiently small volumes.

Building the Fock space of invariant states, we first separate in the fermion 
number. Sates with odd fermion number do not respect the symmetry and we 
can only\Cite{Itoy} have $F=0,2,4$ and 6 (there are six independent Weyl 
components). Particle-hole symmetry relates $F=0$ to $F=6$ and $F=2$ to $F=4$. 
Since $H_f|F=0\ra=H_f|F=6\ra=0$ (the diagonal entries of $\bsigma^i$ 
vanish), this case reduces to the bosonic Hamiltonian which is known not to 
have a zero vacuum energy.\cite{Lue1} So we can restrict our attention to the
$|F=2\ra$ states. The index will be twice the number of zero-energy states
in this sector (due to the particle-hole symmetry, see below). It should be 
noted that only the relative fermion number is well-defined, because 
integrating out certain modes involves filling negative energy (one-particle) 
gluino states. 

\subsection{Invariant two-gluino states}\label{subsec:2gluino}
Instead of making irreducible decompositions of the variables,\cite{Itoy} we 
chose to write down the most general $F=2$ states, and show how the Hamiltonian
acts on these. We can combine two-spinors symmetric or antisymmetric in the 
gauge index (and thus respectively antisymmetric and symmetric in the spinor 
index)
\bea
&&|\cV\ra\equiv{\cV_j}^a{\cI^j}_a\equiv-2i\cV_j^c\veps_{abc}
\blambda^a_\dalpha(\bsigma^{j0})^\dalpha_\dbeta\blambda^{b\dbeta}|0\ra,
\nonumber\\&&|\cS\ra\equiv\cS_{ab}\cJ^{ab}\equiv-\cS_{ab}\blambda^a_\dalpha
\blambda^b_\dbeta\epsilon^{\dalpha\dbeta}|0\ra,\label{eq:SV}
\eea
where $\bsigma^{\mu\nu}=\quart(\bsigma^\mu\sigma^\nu-\bsigma^\nu\sigma^\mu)$,
such that $\bsigma^{j0}=\half\tau_j$ as a $2\times2$ matrix with the first
index up and the second down (bringing also the first index down leads to 
a symmetric $2\times2$ matrix). Here $\cV_j^a$ and $\cS^{ab}=\cS^{ba}$ are
arbitrary (and assumed to depend on $\c_i^a$). The action of the Hamiltonian 
on these states preserves this structure
\be
H_f(|\cS\ra+|\cV\ra)=|\tilde\cV\ra+|\tilde\cS\ra
\ee
with
\bea
&&\tilde{\cV_i}^a=\veps_{ijk}\veps_{abd}\c_j^b\cV_k^d+\c_i^a
\cS^{bb}-\c_i^b\cS^{ba},\nonumber\\ 
&&\tilde\cS^{ab}=2\delta^{ab}\c_i^d\cV_i^d-\c_i^a\cV_i^b-\c_i^b\cV_i^a.
\label{eq:tSV}
\eea
Covariance allows us to write the most general form 
\bea
&&{\cV_j}^a=h_1(\r,u,v)\c_j^a/\r-h_2(\r,u,v)\hat B_j^a/\r^2+h_3(\r,u,v)
\c_j^b\c_k^b\c_k^a/\r^3,\nonumber\\
&&\cS^{ab}=h_4(\r,u,v)\delta^{ab}+h_5(\r,u,v)\c_j^a\c_j^b/\r^2+h_6(\r,u,v)
\c_j^a\c_j^d\c_k^d\c_k^b/\r^4,\label{eq:SVg}
\eea
in terms of the invariants $\r^2=(\c_j^a)^2$, $u=\r^{-4}(\hat B_j^a)^2$ and 
$v=\r^{-3}\det\c$. It is useful to introduce the polar decomposition\Cite{Savv} 
$\c_i^a=\sum_j R_{ij}x_jT^{ja}$, with $R,T\in SO(3)$, since the spherical 
symmetry and gauge invariance allows us to put $\c_i^a=\diag(x_1,x_2,x_3)$. 
Note that this does not completely fix the freedom under gauge and spatial 
rotations. The remnant symmetry involves permutations of the $x_i$ and 
simultaneously flipping two of its signs. In terms of $\vec x$, $\r^2=\sum_j
x_j^2$, $u=\r^{-4}\sum_{i>j}x_i^2x_j^2$ and $v=\r^{-3}\prod_jx_j$, properly 
invariant under the remnant symmetry. Fixing this remnant symmetry, for 
example by $|x_1|\leq x_2\leq x_3$, allows one to solve the $x_i$ from 
$(\r,u,v)$. That $\cV$ and $\cS$ can be expanded each in terms of three 
invariant functions (the $h_m$) is most easily established in this diagonal 
representation 
\bea
&&{\cV_j}^a=\delta_j^a\left(h_1(\r,u,v)\r^{-1}x_j^a+h_2(\r,u,v)\r^{-2}
\det\c/x_j+h_3(\r,u,v)\r^{-3}x_j^3\right),\nonumber\\
&&\cS^{ab}=\delta^{ab}\left(h_4(\r,u,v)+h_5(\r,u,v)\r^{-2}x_a^2+h_6(\r,u,v)
\r^{-4}x_a^4\right),\label{eq:dSV}
\eea
(note that $\hat B_j^a=-\delta_j^ax_j^{-1}\det\c$; no summations over repeated 
indices). Any higher order term can be reduced to this form. This is best 
illustrated by examples: one brings $\delta_j^ax_j^5$ and $\delta^{ab}x_a^6$ 
to the respective form of $\cV^a_j$ and $\cS^{ab}$ in Eq.\Ref{eq:dSV}, by 
using the identities $x_j^5=\r^2x_j^3-\r^4ux_j+\r^6v^2/x_j$ and 
$x_a^6=\r^2x_a^4-\r^4ux_a^2+\r^6v^2$.

We have established that any invariant state $|\Psi\ra$ can be decomposed as
\be
|\Psi\ra=\sum_{m=1}^6h_m(\r,u,v)|e_m(u,v)\ra,
\ee
where the $|e_m\ra$ are implicitly defined by Eqs.~(\ref{eq:SV},\ref{eq:SVg}).
Since $H_f$ does not contain any derivatives with respect to $\c$, we can 
diagonalize $H_f$ pointwize. First determine the matrix of $H_f$ with
respect to the basis $|e_m\ra$
\be
H_f|e_m\ra=\sum_{n=1}^6|e_n\ra H_f^{nm}.
\ee
{}From Eq.\Ref{eq:tSV} one directly reads off that
\be
H_f^{mn}=\r\pmatrix{0&1&-v&2&1&1-u\cr
                    2&0&1&0&0&-v\cr
                    0&-1&0&0&-1&-1\cr
                    2&4v&2-4u&0&0&0\cr
                    -2&0&0&0&0&0\cr
                    0&0&-2&0&0&0\cr}.\label{eq:HF}
\ee
This matrix is not symmetric due to the fact that the $|e_m\ra$ are in general 
not orthogonal. The non-diagonal norm matrix $N^{mn}\equiv\la e_m|e_n\ra$ is 
however block diagonal, since $\la\cS|\cV\ra=0$ for any choice of $\cV$ and 
$\cS$. Introducing the notation $X=1-2u$, $Y=1-3u+3v^2$, $Z=1+2u^2-4u+4v^2$ 
one finds
\be
N^{mn}=\pmatrix{8&24v&8X&0&0&0\cr
                24v&8u&8v&0&0&0\cr
                8X&8v&8Y&0&0&0\cr
                0&0&0&12&4&4X\cr
                0&0&0&4&4X&4Y\cr
                0&0&0&4X&4Y&4Z\cr}.\label{eq:norm}
\ee
The matrix $N$ can be used to make an orthonormal basis. It transforms
$H_f$ to $N^{\half}\cdot H_f\cdot N^{-\half}$. This is seen to be symmetric, 
e.g. by establishing the symmetry of $N\cdot H_f$.

\begin{figure}[htb]
\vspace{4.5cm}
\includegraphics{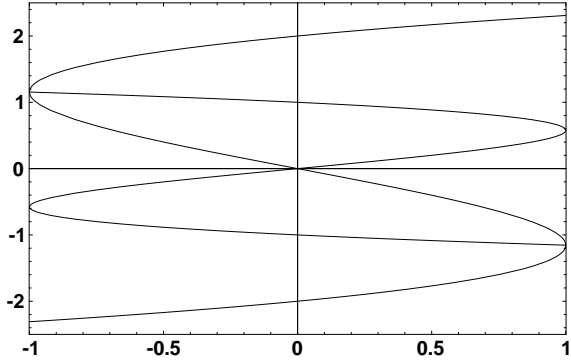}
\caption{The six eigenvalues $E_f/\r$, of $H_f/\r$, plotted versus 
$3v\protect\sqrt{3}$.}\label{fig:EFs}
\end{figure}

We denote by $E_f$ the energies of the two-gluino state in a given background. 
With $E_f$ linear in $\r$ and, as we will see, {\em independent} of $u$, 
only the $v$ dependence is non-trivial. The eigenvalues of $H_f$, for the six 
invariant two-gluino states, are determined by 
\be
\det(H_f-E_f)=(E_f^3-4\c^2E_f-16\det\c)(E_f^3-\c^2E_f+2\det\c)=0,
\label{eq:eigen}
\ee
In terms of the three roots $\mu_i$, $E_f^3-\c^2E_f+2\det\c=(E_f-\mu_1)
(E_f- \mu_2)(E_f-\mu_3)$, the remaining three roots are given by 
$\tilde\mu_i= -2\mu_i$. Note that $\sum_i\mu_i=0$, $\sum_i\mu_i^2=2\c^2$ 
and $\prod_i\mu_i=-2\det\c$. Explicit expressions for the two-gluino 
energies are (see Fig.~\ref{fig:EFs})
\be
\tilde\mu_k=-2\mu_k=-4\r\cos(2\pi k/3+\arccos[-3v\sqrt{3}]/3)/\sqrt{3},
\label{eq:mu}
\ee
with $\arccos$ ranging from 0 to $\pi$. Note that $3v\sqrt{3}\in[-1,1]$ since 
$\prod x_i$ is extremal (under the constraint $\sum_ix_i^2=1$) when all $|x_i|$
are equal (to $1/\sqrt{3}$). The lowest eigenvalue, $E^{\rm min}_f\equiv\tilde
\mu_3$, takes on the value $-2\r$ along the vacuum valley, since $u=0$ implies 
$v=0$. The eigenvectors associated to the eigenvalue $\tilde\mu_k$ are also 
easily constructed, 
\be
\tilde h^{(k)}=\left(\half\tilde\mu_k/\r,1,0,1+4v\r/\tilde\mu_k,-1,0\right)/
|4(1+6v\r/\tilde\mu_k)|,\label{eq:gs}
\ee
normalized {\em in a fixed background} according to $\sum_{m,n=1}^6
\tilde h_m^{(k)}N^{mn}\tilde h^{(k')}_n=\delta_{kk'}$. We will not write 
down the more complicated analytic form for the eigen functions belonging 
to the remaining three eigenvalues. Only $\tilde h^{(3)}$ will be referred
to later when discussing projection to the vacuum valley.

At this point it is perhaps interesting to observe that along the vacuum 
valley (two out of the three $x_j$ vanish), the particle-hole dual of 
$|\Psi\ra=\tilde h^{(3)}_m|e_m\ra$ can be expressed as $|\overbar{\Psi}\ra=
\half\blambda_\dalpha^a\c_i^a\c_i^b\blambda_b^\dalpha|\Psi\ra$, obtained 
from $|\Psi\ra$ by creating a pair of abelian zero-momentum gluinos, in 
accordance with the analysis of Witten.\cite{WiIn} To prove this, note 
that with $\c_i^a$ abelian, defining $\cV_i^a=-\c_i^a$ and $\cS_{ab}=
\delta_{ab}-\c_i^a\c_i^b$, we can write $|\Psi\ra=|\cV\ra+|\cS\ra$. Its 
particle-hole dual is given by $|\overbar{\Psi}\ra=|\overbar{\cV}\ra+
|\overbar{\cS}\ra$, with 
\bea
&&|\overbar{\cV}\ra\equiv{\cV_j}^a{\overbar{\cI}^j}_a\equiv-2i\cV_j^c\veps_{abc}
\lambda^a_\alpha(\sigma^{j0})^\alpha_\beta\lambda^{b\beta}|\overbar{0}\ra,
\nonumber\\&&|\overbar{\cS}\ra\equiv\cS_{ab}\c\overbar{J}^{ab}\equiv-\cS_{ab}
\lambda^a_\alpha\lambda^b_\beta\epsilon^{\alpha\beta}|\overbar{0}\ra,\quad
|\overbar{0}\ra\equiv\prod_{a\dalpha}\blambda_\dalpha^a|0\ra,\label{eq:bSV}
\eea
also defining the particle-hole symmetry in the general case.

It should be noted that our change of variables from $\c$ to $x_i$ will suffer
from coordinate singularities, also affecting the linear independence of 
the $|e_m\ra$. This is most obvious from the fact that $\det N=32768 J^4$, 
with $J$ (up to a constant) the Jacobian for the change of variables,
$d^9\c=\twoth\pi^4 J d^3x$ (with $|x_1|\leq x_2\leq x_3$),
\be
J\equiv \prod_{i>j}^{\vphantom{j}}|x_i^2-x_j^2|=\r^6\sqrt{u^2(1-4u)-
v^2(4-18u+27v^2)}.\label{eq:jac}
\ee 
Indeed, the normalization of $\tilde h^{(k)}$ in Eq.\Ref{eq:gs} is singular 
when $\tilde\mu_k+6v\r$ vanishes, which is easily identified with a vanishing 
Jacobian. For this observe that $3\c^2E_f+18\det\c=3\r^2(E_f+6v\r)$ equals the 
difference of $E_f^3-4\c^2E_f-16\det\c$ and $E_f^3-\c^2E_f+2\det\c$. Since each
of these vanish for either $E_f=\tilde\mu_i$ or $E_f=\mu_i$, we conclude that 
$\tilde\mu_k+6v\r$ vanishes if and only if $\tilde\mu_k$ coincides with one of 
the other roots $\mu_i$. From Fig.~\ref{fig:EFs} we read off this happens for 
$v=\pm\sqrt{3}/9$, which is the extremal value of $v$ where all three $|x_i|$ 
are equal (constraining $u$ to be $3|v|^{4/3}=1/3$), and therefore $J=0$. As 
long as $J\neq0$, the limit $v\rightarrow0$ exists, e.g.
\be
\lim_{v\rightarrow0}\pmatrix{\tilde h^{(1)}\cr\tilde h^{(2)}\cr\tilde h^{(3)}
   \cr}=\pmatrix{\quart&\quart&0&\quart&-\quart&0\cr0&\half&0&0&-\half&0\cr
                 -\quart&\quart&0&\quart&-\quart&0\cr},\label{eq:hs}
\ee
independent of $u$, but we have to remember that when $J=0$ (in particular
$u=0$) the $|e_m\ra$ are no longer independent. We come back to this when 
projecting to the vacuum valley.

Before constructing a full invariant basis, to gain more confidence that the 
$|e_m\ra$ correctly describe the invariant two-gluino states, it is useful to 
note that the spectrum of $H_f$ can also be obtained from its (non-invariant) 
one-particle states.\cite{Halp} For this we write
\be
H_f=\blambda^a_\dalpha M^{\dalpha\beta}_{ab}\lambda^b_\beta,\quad 
M^{\dalpha\beta}_{ab}=-i\veps_{abd}(\bsigma^j)^{\dalpha\beta}\c_j^d.
\ee
The one-particle fermion energies are given by the eigenvalues of $M$. 
Note that charge conjugation symmetry, $M^*=\tau_2M\tau_2$, implies each
eigenvalue is two-fold degenerate. By {\em direct} computation one verifies
that\Cite{Halp}
\be
M^3-\c^2 M-2\det\c=0,
\ee
such that any eigenvalue $\mu$ of $M$ satisfies the equation $\mu^3-\c^2\mu-2
\det\c=(\mu+\mu_1)(\mu+\mu_2)(\mu+\mu_3)=0$, with $\mu_i$ as defined before. 
Hence the one-particle fermion energies are given by $-\mu_i$, and each occurs
with a two-fold (spin) degeneracy. We can make invariant two-particle states
only by combining two opposite spin states. There are six such invariant 
states, three with the {\em same} one-particle energies, giving $E_f=-2\mu_i=
\tilde\mu_i$, and three with {\em different} one-particle energies, giving 
$E_f=-\mu_i-\mu_j=\mu_k$ with $i\neq j\neq k$. This agrees with our earlier 
results. Note that there are also three triplet two-gluino states with 
two-particle energies $E_f=\mu_k$, combining two one-particle states with 
different energies but equal spin, which are not invariant and thus left out 
from our considerations. See the Appendix for yet another method.

\subsection{Supersymmetric spherical harmonics}\label{subsec:Ys}
We have removed the angular degrees of freedom on which the invariant wave 
functions do not depend: those associated with the gauge and space rotations.
This leaves a six component $F=2$ wave function, depending on either $(\r,u,v)$
or equivalently $(x_1,x_2,x_3)$. It is well known\Cite{Savv,AnnP} that the 
kinetic term of the Hamiltonian reduces to  
\be
-\half\frac{\partial^2}{(\partial\c_i^a)^2}=-\half J^{-1}(\vec x)
\frac{\partial}{\partial x_j}J(\vec x)\frac{\partial}{\partial x_j}=
-\half\left(\r^{-8}\frac{\partial}{\partial\r}\r^8\frac{\partial}{\partial\r}
+\frac{\Delta(u,v)}{\r^2}\right),\label{eq:kin}
\ee
with $\Delta(u,v)$ the Laplacian for $S^8$, acting on {\em invariant} 
functions, given by\Cite{AnnP}
\bea
\Delta(u,v)&=&4(3v^2+u-4u^2)\frac{\partial^2}{(\partial u)^2}
              +8(1-3u)v\frac{\partial^2}{\partial u\partial v}
              +(u-9v^2)\frac{\partial^2}{(\partial v)^2}\nonumber\\
            &&+4(2-11u)\frac{\partial}{\partial u}
              -30v\frac{\partial}{\partial v}.\label{eq:del}
\eea
Note that for the bosonic theory ($F=0$) the variables $(\r,u,v^2)$ were used
(for the positive parity states), but for the $F=2$ sector functions even and 
odd in $v$ mix, requiring us to use the variables $(\r,u,v)$. The Hamiltonian 
still allows for a radial decomposition (note that $\partial_\r|e_m\ra=0$) 
\be
\hat H^{mn}=-\half\delta^{mn}\r^{-8}\frac{\partial}{\partial\r}\r^8
             \frac{\partial}{\partial\r}+\half\delta^{mn}\r^4u+\r
             \hat H_f^{mn}+\r^{-2}\hat H_\Delta^{mn},\label{eq:Hop}
\ee
where $\hat H_f^{mn}\equiv H_f^{mn}/\r$, see Eq.\Ref{eq:HF}, and
\be
\sum_{m,n=1}^6|e_n(u,v)\ra\hat H_\Delta^{nm}h_m(\r,u,v)
=-\half\Delta(u,v)\sum_{m=1}^6h_m(\r,u,v)|e_m(u,v)\ra.
\ee

It is straightforward, but quite tedious, to explicitly calculate 
$\hat H_\Delta^{mn}$, 
\be
\hat H^{mn}_\Delta=-\half\delta^{mn}\Delta(u,v)-\half\pmatrix{
                    \Delta^1_\cV&\oslash\cr\oslash&\Delta^1_\cS\cr}
                    -\half\pmatrix{\Delta^0_\cV&\oslash\cr
                                   \oslash&\Delta^0_\cS\cr}.\label{eq:Hdel}
\ee
For the first term $\Delta$ only acts on the coefficient functions, for the
next we collect the terms where one derivative acts on the coefficient
functions and one on the two-gluino basis vectors $|e_m\ra$, whereas the
last term has all derivatives acting on these basis vectors. This splits 
in the two sectors associated to $|\cV\ra$ and $|\cS\ra$, since derivatives 
cannot lead to mixing of these two-gluino states, specified by $\cI$ and 
$\cJ$ in Eq.\Ref{eq:SV}. We find the following results
\be
\Delta^1_\cV\equiv2\pmatrix{(2-4u)\partial_u-3v\partial_v&\partial_v+
                       2v\partial_u&3v\partial_v+6u\partial_u\cr
       \partial_v&(2-8u)\partial_u-6v\partial_v&-6v\partial_u\cr
         -2\partial_u&-\partial_v&-12u\partial_u-9v\partial_v\cr},
\ee
\be
\Delta^1_\cS\equiv2\pmatrix{0&2v\partial_v&-8v^2\partial_u\cr
    0&(4-8u)\partial_u-6v\partial_v&4v\partial_v+8u\partial_u\cr
                   0&-4\partial_u&-16u\partial_u-12v\partial_v\cr},
\ee
and
\be
\Delta^0_\cV\equiv2\pmatrix{\!-4&~0&~7\cr~0&\!-9&~0\cr~0&~0&\!-15\cr},\quad
\Delta^0_\cS\equiv\pmatrix{~0&~3&~1\cr~0&\!-9&~11\cr~0&~0&\!-22\cr}.
\ee

Invariant wave functions, $|\Psi\ra=\sum_{m=1}^6h_m(\r,u,v)|e_m(u,v)\ra$, are 
normalized using
\be
\la\Psi|\Psi^\prime\ra=\int d^9\c\sum_{m,n=1}^6h^*_mN^{mn}h^\prime_n,
\ee
with $N^{mn}$ the norm matrix as defined in Eq.\Ref{eq:norm}. The 
matrix elements of the Hamiltonian with respect to such a basis are
thus given by
\be
\la\Psi|H|\Psi^\prime\ra=\int d^9\c\sum_{m,n,p=1}^6h^*_mN^{mn}\hat H^{np}
h^\prime_p,
\ee
with $\hat H^{mn}$ the matrix operator as defined in Eq.\Ref{eq:Hop}.

Like for the bosonic case (with $\hat H_\Delta\equiv-\half\Delta$ and $\hat 
H_f\equiv0$) we first construct an invariant set of ``angular" wave functions 
as polynomials in $u$ and $v$. These can be chosen to be eigenfunctions 
($\cY_s^{xyz}$) of $\hat H_\Delta$, proportional to coefficient functions 
$U_s^z$ defined by (for $F=0:\,z=0;\,s\in\{2,5\}$ and $F=2:\,z\in\{1,2,3\};$
$s\in\{0,1,2,3\}$) 
\bea
\pmatrix{U^1_0\cr U^2_0\cr U^3_0\cr}=\pmatrix{0&0&0&1&0&0\cr
                                0&0&0&0&1&0\cr0&0&0&0&0&1\cr}&,& 
\pmatrix{U^1_1\cr U^2_1\cr U^3_1\cr}=\pmatrix{1&0&0&0&0&0\cr
                                0&0&1&0&0&0\cr0&v&0&0&0&0\cr},\nonumber\\
\pmatrix{U^1_2\cr U^2_2\cr U^3_2\cr}=\pmatrix{0&1&0&0&0&0\cr
                                v&0&0&0&0&0\cr0&0&v&0&0&0\cr}&,&
\quad U^z_3=vU^z_0,\quad U_5^0=vU_2^0=v,\label{eq:Us}
\eea
{\em judiciously} chosen such that $q=2x+3y+z$ allows us to order states, and
\be
\hat H_\Delta u^xv^{2y}U_s^z=\half(2q+s-2)(2q+s+5)u^xv^{2y}U_s^z+\delta 
R_s^{xyz},
\label{eq:mon}
\ee
with $\delta R_s^{xyz}$ a linear combination of these monomials, but of {\em 
lower order} in $q$ (states with different $s$ do not mix under $\hat H_\Delta$,
but they {\em do} mix under $\hat H_f$). The appearance of the combination $2x
+3y$ is related to the fact that $u^xv^{2y}=r^{-2(2x+3y)}(\hat B^2)^x(\det 
\c)^{2y}$. The eigenvalues of $\hat H_\Delta$ can be written as $L(2L+7)$, 
with the ``angular momentum" $L=q+s/2-1=2x+3y+z+s/2-1$ taking half integer 
values. Eq.\Ref{eq:mon} allows us to solve $[\hat H_\Delta-L(2L+7)]\cY_s^{xyz}
=0$ for $R_s^{xyz}$ (of order $q<2x+3y+z$), with
\be
\cY_s^{xyz}\equiv u^xv^{2y}U_s^z+R_s^{xyz}.
\ee

\begin{table}[htb]
\caption{All orthonormal spherical harmonics for $L<4$, with
$L=2x+3y+z+s/2-1$, such that $\hat H_\Delta\hat\cY_s^{xyz}=L(2L+7)
\hat\cY_s^{xyz}$. For higher $L$ use the available program.\protect\cite{Code}}
\label{tab:harm}
\vspace{0.2cm}
\begin{center}
\begin{tabular}{|c|l|}
\hline
$L$&$\hat\cY_{n\hph0}\!=\hat\cY_s^{xyz}\!=(h_1,h_2,h_3,h_4,h_5,h_6)$\\
\hline
$0$&$\hat\cY_{1\hph0}=\hat\cY_0^{001}=(0,0,0,1,0,0)
            \sqrt{35/2}/8\pi^2$\\ 
$\half$&$\hat\cY_{2\hph0}=\hat\cY_1^{001}=(1,0,0,0,0,0)
            \sqrt{105}/16\pi^2$\\ 
$1$&$\hat\cY_{3\hph0}=\hat\cY_2^{001}=(0,1,0,0,0,0)
            \sqrt{1155/2}/16\pi^2$\\ 
$1$&$\hat\cY_{4\hph0}=\hat\cY_0^{002}=(0,0,0,-\frac{1}{3},1,0)
            3\sqrt{77}/16\pi^2$\\ 
$\thalf$&$\hat\cY_{5\hph0}=\hat\cY_3^{001}=(0,0,0,v,0,0)
            \sqrt{15015}/16\pi^2$\\ 
$\thalf$&$\hat\cY_{6\hph0}=\hat\cY_1^{002}=(-\frac{7}{11},0,1,0,0,0)
            11\sqrt{273/5}/32\pi^2$\\ 
$2$&$\hat\cY_{7\hph0}=\hat\cY_2^{002}=(v,-\frac{1}{13},0,0,0,0)
            39\sqrt{77}/32\pi^2$\\ 
$2$&$\hat\cY_{8\hph0}=\hat\cY_0^{003}=(0,0,0,\frac{10}{143},-\frac{11}{13},1)
            429\sqrt{7/86}/16\pi^2$\\ 
$2$&$\hat\cY_{9\hph0}=\hat\cY_0^{101}=(0,0,0,-\frac{6}{43}+u,-\frac{22}{43},
            \frac{26}{43})3\sqrt{6149/2}/16\pi^2$\\ 
$\fhalf$&$\hat\cY_{10}=\hat\cY_3^{002}=(0,0,0,-\frac{v}{3},v,0)
            3\sqrt{51051/2}/16\pi^2$\\ 
$\fhalf$&$\hat\cY_{11}=\hat\cY_1^{003}=(-\frac{11}{195},v,\frac{1}{15},
               0,0,0)39\sqrt{1785}/64\pi^2$\\ 
$\fhalf$&$\hat\cY_{12}=\hat\cY_1^{101}=(-\frac{1}{4}+u,-\frac{13v}{44},
               \frac{5}{44},0,0,0)33\sqrt{221/7}/16\pi^2$\\ 
$3$&$\hat\cY_{13}=\hat\cY_2^{003}=(-\frac{10v}{17},\frac{1}{51},v,0,
               0,0)51\sqrt{4389/5}/32\pi^2$\\ 
$3$&$\hat\cY_{14}=\hat\cY_2^{101}=(-\frac{6v}{13},-\frac{12}{65}+u,
               \frac{38v}{65},0,0,0)39\sqrt{17765/7}/64\pi^2$\\ 
$3$&$\hat\cY_{15}=\hat\cY_0^{011}=(0,0,0,\frac{4}{663}-\frac{u}{17}+
               v^2,0,0)663\sqrt{209/7}/32\pi^2$\\ 
$3$&$\hat\cY_{16}=\hat\cY_0^{102}=(0,0,0,\frac{2}{51}-\frac{3u}{17},
               -\frac{6}{17}+u,\frac{4}{17})51\sqrt{2717/7}/32\pi^2$\\ 
$\shalf$&$\hat\cY_{17}=\hat\cY_3^{003}=(0,0,0,\frac{28v}{323},
               -\frac{15v}{19},v)969\sqrt{231/10}/32\pi^2$\\ 
$\shalf$&$\hat\cY_{18}=\hat\cY_3^{101}=(0,0,0,-\frac{12v}{65}+uv,
              -\frac{6v}{13},\frac{38v}{65})39\sqrt{53295/2}/32\pi^2$\\ 
$\shalf$&$\hat\cY_{19}=\hat\cY_1^{011}=(\frac{10}{969}-\frac{u}{19}+
               v^2,-\frac{2v}{19},-\frac{2}{323},0,0,0)969\sqrt{429/14}/32\pi^2
               $\\ 
$\shalf$&$\hat\cY_{20}=\hat\cY_1^{102}=(\frac{44}{399}-\frac{93u}{133}+
               \frac{2v^2}{7},\frac{2v}{7},-\frac{20}{133}+u,0,0,0)
               19\sqrt{51051/2}/64\pi^2$\\
\hline
\end{tabular}
\end{center}
\end{table}

To order these states at degenerate values of $L$, we begin at $L=0$ with 
$\cY_1=\cY_0^{001}$. Raising $L$ with $\half$ we start with the lowest 
value of $q$ and construct the independent, but in general non-orthogonal 
states $\cY_s^{xyz}$. Using that $y$ and $z$ are uniquely fixed by $q-2x$, we 
start with $x=0$ and every time we increase $x$ by one, we modify $\cY_s^{xyz}$
by projecting it on the orthogonal complement of the previous states (the 
Gramm-Schmidt procedure). When completed we increase $q$ until all states with 
a given value of $L$ are constructed, after which we increase $L$. Dividing by 
the norm we thus obtain a complete orthonormal set of ``spherical harmonics", 
$\hat\cY_n(u,v)\equiv\la u,v|n\ra$, with $n$ labelling the $n$-th state thus 
constructed. The first few spherical harmonics $\hat\cY_n(u,v)$ are collected 
in Table~\ref{tab:harm}. Note that to evaluate inner products we need to 
compute the integrals $X_{x,y}\equiv\int_{\r=1}d^9\c~u^xv^{2y}$. For this we 
recall the recursive definition\Cite{AnnP}
\be
X_{i,j}=\frac{4i(1+i+4j)X_{i-1,j}+12i(i-1)X_{i-2,j+1}+2j(2j-1)X_{i+1,j-1}}{
                             (4i+6j)(4i+6j+7)},
\ee
with $X_{0,0}=32\pi^4/105$. The Mathematica\Cite{Math} code for generating the 
$\hat\cY_n$ is available through the World Wide Web.\cite{Code} 

\begin{figure}[htb]
\vspace{7.8cm}
\includegraphics{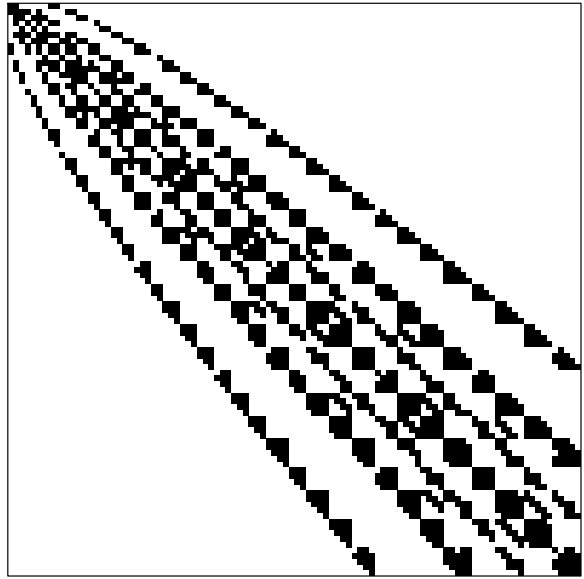}
\caption{Band structure of the reduced Hamiltonian, for the first 100 states.}
\label{fig:band}
\end{figure}

We denote by $L_n$ the value of $L=2x+3y+z+s/2-1$ implicitly defined by $\hat
\cY_n=\hat\cY_s^{xyz}$. It will be convenient to also introduce $\ell\equiv 
2L+3$. With the spherical harmonics we can now construct a reduced Hamiltonian,
\bea
&&\la n'|H|n\ra=H^{n'n}(\r)=K(\r;L_n)\delta^{n'n}+\half\r^4\la n'|u|n\ra+
                                       \r\la n'|\hat H_f|n\ra,\nonumber\\
&&K(\r;L)=-\half\r^{-8}\partial_\r\r^{8}\partial_\r+\r^{-2}L(2L+7)=
                                       \r^{-3}\K(\r,\ell)\r^3,\nonumber\\
&&\K(\r;\ell)=-\half\r^{-2}\partial_\r\r^{2}\partial_\r+
                                  \half\r^{-2}\ell(\ell+1).\label{eq:Hred}
\eea
We illustrate its sparse nature by showing in Fig.~\ref{fig:band} the entries 
where either $\la n'|u|n\ra\neq0$ or $\la n'|\hat H_f|n\ra\neq0$ as black 
squares. The different bands can be traced to come from the selection rules 
$|\delta L|=0,1,2$ for the matrix elements of $u$, and $|\delta L|=\half$ for 
the matrix elements of $\hat H_f$. The number of codiagonals is bound by 
$-3+4\sqrt{n}$. 

To write down the matrix of the full Hamiltonian with respect to an invariant 
basis, we introduce radial wave functions $\phi_p^\ell(\r)\equiv\la\r|p,\ell
\ra$. The radial quantum number $p$ is associated to a momentum if we choose
\be
\phi_p^\ell(\r)=C_p^\ell\r^{-3}j_\ell(k^\ell_p\r),\quad
E_p^\ell=\half(k_p^\ell)^2,\label{eq:besj}
\ee
where $j_\ell(z)$ is the spherical Bessel function which satisfies the equation
$\K(\r;\ell)j_\ell(\r)=\half j_\ell(\r)$ or $K(\r;L)\phi_p^{2L+3}(\r)=
E_p^{2L+3}\phi_p^{2L+3}(\r)$. Normalization with a factor $C_p^\ell$ is such 
that $\int d\r~\r^8\phi_p^\ell(\r)^*\phi_q^\ell(\r)=\delta_{pq}$. In terms 
of the normalized basis of invariant states $\la\c|p,n\ra\equiv\la u,v|n\ra
\la\r|p,\ell_n\ra$, the matrix of the Hamiltonian is given by $H_{n'n}^{p'p}=
\la p',n|H|p,n\ra$, or
\be
H_{n'n}^{p'p}=E_p^{\ell_n}\delta_{nn'}\delta^{pp'}+
              \la p',\ell_{n'}|\r^4|p,\ell_n\ra\la n'|\frac{u}{2}|n\ra+
              \la p',\ell_{n'}|\r|p,\ell_n\ra\la n'|\hat H_f|n\ra.
\ee
To complete the construction of the basis, we need to address the question of 
boundary conditions (fixing the momenta), which somehow have to incorporate 
that the groundstate wave function becomes a constant, after projecting to 
the vacuum valley.

\section{Vacuum valley and boundary condition}\label{sec:val}
The vacuum valley is characterized by those configurations for which $u=0$
(this implies $v=0$), and $\r$ measures the (three-dimensional) distance to 
the origin along this vacuum valley. The wave function can be decomposed as 
$\Psi=\sum_{n=0}^\infty\r^{-3}f_n(\r)\chi_n(u,v;\r)$, with $\chi_n(u,v;\r)$ 
normalized eigenfunctions of $\H(u,v;\r)=\hat H(u,v,\r)|_{\partial_\r=0}$,
see Eq.\Ref{eq:Hop}. It is convenient to use the spherical 
coordinates 
\bea
&&\vec x=\r\left(\sin(\vphi)\sin(\theta),\cos(\vphi)\sin(\theta),\cos(\theta)
\right),\\&&u=\sin^2\theta(\cos^2\theta+\cos^2\vphi\sin^2\vphi\sin^2\theta),
\quad v=\sin^2\theta\cos^2\theta\cos\vphi\sin\vphi,\nonumber\\&&J=\r^6|\J|,
\quad\J=\sin^2\theta\cos(2\vphi)\left(1-3\sin^2\theta+\sin^4\theta[17-\cos(
4\vphi)]/8\right),\nonumber
\eea
With $\chi_n(\theta,\vphi;\r)\equiv\chi_n(u(\theta,\vphi),v(\theta,\vphi);\r)$,
we define
\bea
&&\la\chi|\chi'\ra=\frac{\pi^3}{6}\int_{-\pi/4}^{\pi/4} d\vphi\int_0^{\theta(
\vphi)}d\theta \sin\theta\J(\theta,\vphi)\chi^*(\theta,\vphi;\r)\chi'(\theta,
\vphi;\r),\nonumber\\&&\la\Psi|\Psi'\ra=\sum_{n=0}^\infty\int 4\pi\r^2d\r 
f_n^*(\r)f^\prime_n(\r),\label{eq:inchi}
\eea
where $\theta(\vphi)$ implements the constraint $x_2\leq x_3$ (we can also 
take $\theta\in[0,\pi]$ and $\vphi\in[0,2\pi]$, in which case one needs to 
replace $\J$ with $|\J|/24$). We moved a factor of $4\pi$ to the measure for 
the $\r$ integration, such that $f(\r)$ can be interpreted as the vacuum-valley
wave function (in the S-wave channel). The Hamiltonian reduces to
\bea
&&H\Psi=\sum_{n,k,m}\chi_n\left(-\half\r^{-2} D^{nm}_\r(\r^2 D^{mk}_\r)+
\delta^{nk}(E_n(\r)+6\r^{-2})\right)f_k(\r),\nonumber\\ &&D^{mn}_\r=\delta^{mn}
\partial_\r+A^{mn}_\r(\r),\quad A^{mn}_\r(\r)=\la\chi_m|\partial_\r|\chi_n\ra,
\label{eq:berA}
\eea
with $E_n(\r)$ the eigenvalues of $\H$, i.e. $\H\chi_n=E_n(\r)\chi_n$. The 
pairs $E_n(\r)$, $\chi_n/\sqrt{4\pi}$ can be approximated by the eigenvalues 
and normalized eigenvectors of the truncated reduced Hamiltonian, replacing 
$K(\r;L_n)$ with $\r^{-2}L_n(2L_n+7)$ in Eq.\Ref{eq:Hred}. From the numerical
point of view, for large $\r$ the truncation starts to become a problem due 
to the progressive localization of $\chi_0$, requiring ever larger values of 
$L$. With $L<20$ we can get accurate results for $\r$ up to 5. As a check, 
we will also expand $\r^{-1}E_0(\r)$ and $\chi_0(\theta,\vphi;\r)$ in powers 
of $\r^{-3}$. In the adiabatic (or Born-Oppenheimer) approximation which 
neglects $f_{n>0}$, Eq.\Ref{eq:berA} becomes
\bea
&&H\Psi=\chi_0\left(-\half\r^{-2}\partial_\r\r^2\partial_\r+V_\eff(\r)\right)
f_0(\r),\nonumber\\&&V_\eff(\r)=\half A^2(\r)+E_0(\r)+6\r^{-2},\label{eq:Veff}
\eea
with $A^2(\r)\equiv-\sum_n A_\r^{0n}(\r)A_\r^{n0}(\r)=\la\partial_\r\chi_0|
\partial_\r\chi_0\ra$. Unbroken supersymmetry requires $V_{\eff}(\r)=0$. 

An asymptotic analysis\Cite{LNdW,Halp,Froh} is complicated by the coordinate 
singularities. The transverse coordinate is $\theta$, with $\exp(-\half
\theta^2\r^3)$ the leading exponential factor in $\chi_0$. Introducing 
$\hat\theta=\theta/\r^{3/2}$, we can expand $\H$ 
\bea
&&\r^{-1}\H^{mn}=\half\delta^{mn}\hat\theta^2+\hat H_f^{mn}(0)-\half
\delta^{mn}\hat\Delta-\half\pmatrix{\hat\Delta_\cV&\oslash\cr\oslash&
\hat\Delta_\cS\cr}+\cO(\r^{-3}),\nonumber\\ 
&&\hat\Delta_\cV\equiv 2\pmatrix{\hph0\hat\partial_1&\hph0\hat\partial_2&0\cr
                               \hph0\hat\partial_2&\hph0\hat\partial_1&0\cr
                                   -\hat\partial_1&-\hat\partial_2&0\cr},\quad
\hat\Delta_\cS\equiv 4\pmatrix{0&0&0\cr 0&\hph0\hat\partial_1&0\cr
                               0&-\hat\partial_1&0\cr},\nonumber\\ 
&&\hat\partial_1\equiv\hat\theta^{-1}\partial_{\hat\theta}-\hat\theta^{-2}
\tan(2\vphi)\partial_\vphi,\quad\hat\partial_2\equiv\hat\theta^{-2}\sec(2\phi)
\partial_\vphi,\nonumber\\
&&\hat\Delta\equiv\hat\theta^{-3}\partial_{\hat\theta}\hat\theta^3
                  \partial_{\hat\theta}+\hat\theta^{-2}\sec(2\vphi)
                  \partial_\vphi\cos(2\vphi)\partial_\vphi.\label{eq:Hopa}
\eea
There is one exact zero-energy state, $\hat\chi_0=h\exp(-\half\hat\theta^2)$,
with the proper normalization conveniently expressed as
\be
\chi_0=(1+\cO(\r^{-3}))h\frac{\r^3\sqrt{3}}{2(\sqrt{\pi})^3}\exp(-\half u\r^3),
\quad h=(2,-2,2,-4,1,3)/16.\label{eq:chi0}
\ee
To first order we find $\la\partial_\r\chi_0|\partial_\r\chi_0\ra=9/(2\r^2)$ 
and $E_0(\r)=-33/(2\r)^2$, which indeed gives $V_{\eff}(\r)=0+\cO(\r^{-5})$. 
Note that $h\neq\tilde h^{(3)}$, but still $h N(0) \hat H_f(0) h=-2h N(0)h=-2$ 
(see Eqs.~(\ref{eq:norm},\ref{eq:hs})). This can occur because $h$ and 
$\tilde h^{(3)}$ differ by an element from the (three dimensional) kernel of 
$N(0)$. We used $\H$ to find $\chi_0$, as opposed to $N(0)\H$, to 
guarantee $\la\chi|\H|\chi_0\ra$ vanishes to lowest order for {\em any} 
function $\chi(u,v)$. The non-trivial kernel makes it particularly cumbersome 
to perform the asymptotic expansion. 

\subsection{Numerical results for $V_\eff$}\label{subsec:Vnum}
\begin{figure}[htb]
\vspace{6.7cm}
\includegraphics{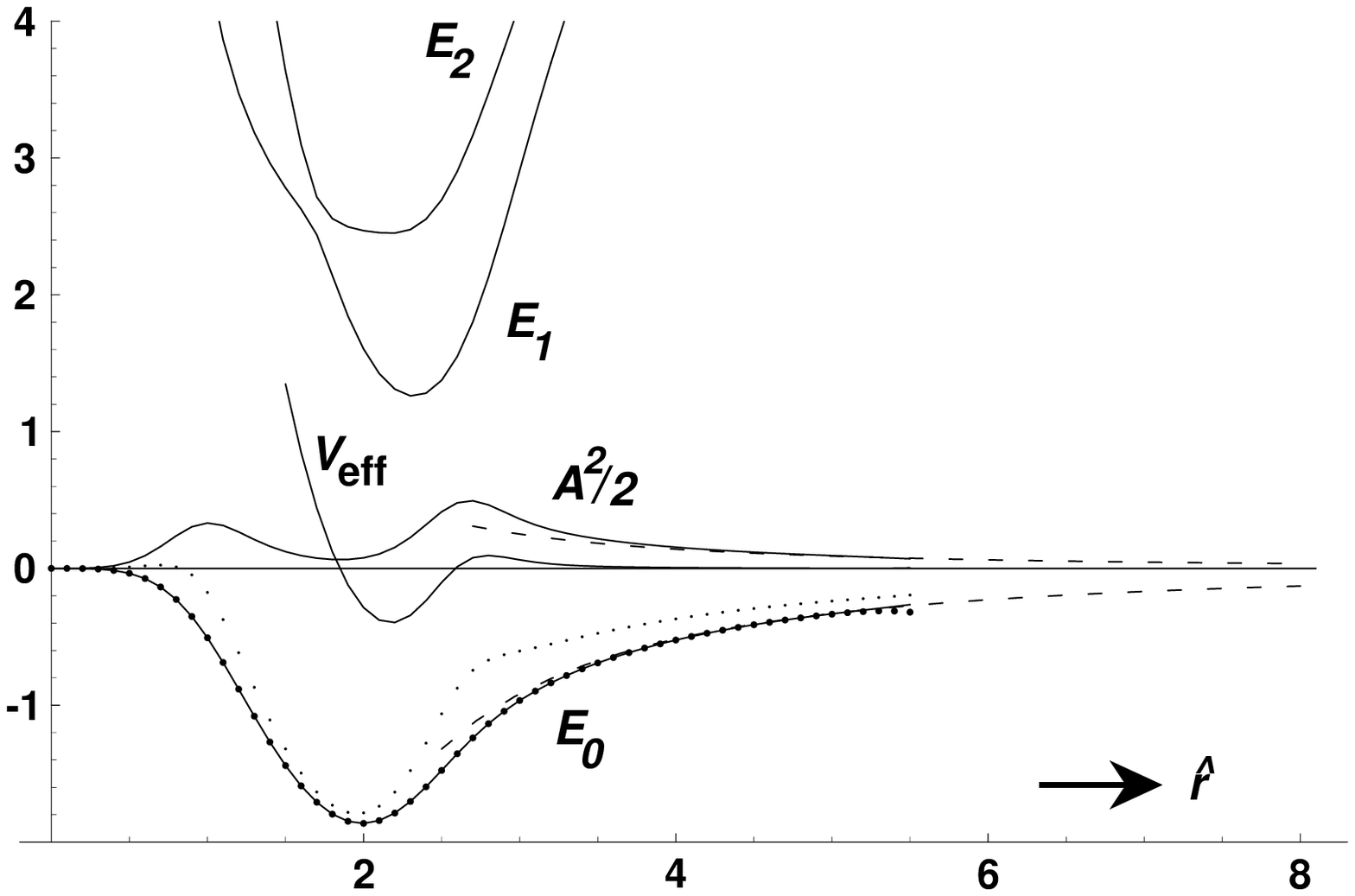}
\includegraphics{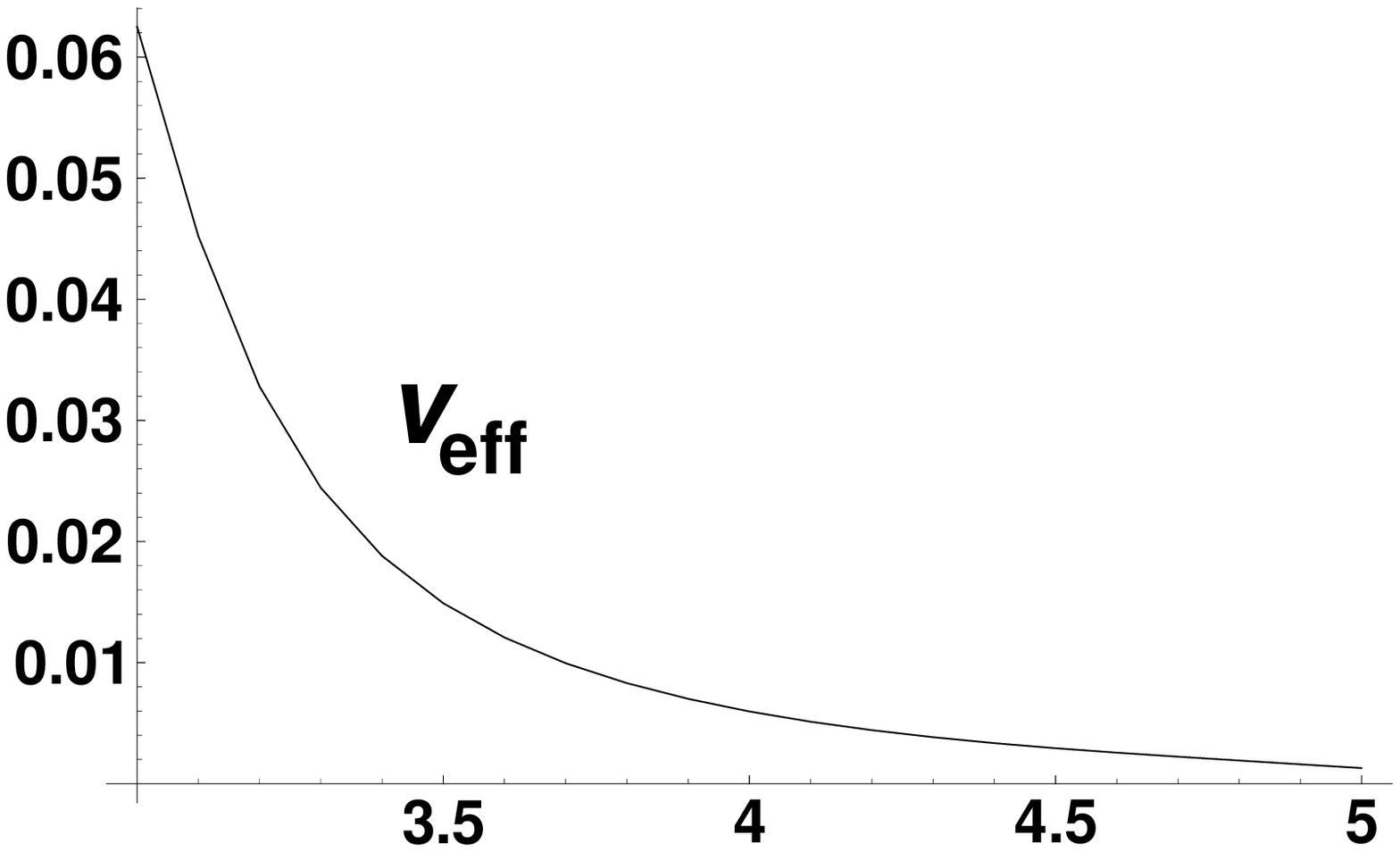}
\caption{The energies $E_n(\r)$, for $n=0,1,2$, $\half A^2(\r)\equiv\half
\la\partial_\r\chi_0|\partial_\r\chi_0\ra$ and the vacuum valley effective
potential $V_\eff(\r)$ (see also the inset). The dashed lines give 
asymptotic results. The dotted curve shows $V_\eff(\r)-6\r^{-2}=E_0(\r)
+ \half A^2(\r)$. The (larger) dots represent the lower bound on $E_0$, 
using Temple's inequality.}
\label{fig:Ers}
\end{figure}

Much more convincing, however, are the numerical results depicted in 
Fig.~\ref{fig:Ers}, as these ``sum" to all orders in $\r^{-3}$. We truncated
the matrix $\la n'|\H|n\ra$ to the first 420 spherical harmonics $\hat\cY_n$, 
whose eigenvalues approximate $E_n$. Temple's inequality,\cite{SiRe,Kovb}
 $0\leq 
\la\H\ra-E_0\leq\la\,[\,\H-\la\H\ra\,]^2\,\ra/
\Delta E$, can be used to bound the error due to the truncation, with 
$\Delta E=E_1-E_0$ giving a safe bound. In Fig.~\ref{fig:Ers} this lower 
bound is indicated by the dots, and the upper bound by the drawn lines. 
The dashed curves ($-33/(2\r)^2$ and $9/(2\r)^2$) demonstrate that our 
numerical results perfectly match the asymptotic expansion derived above.

We also show $E_1(\r)$, to illustrate the gap in the transverse fluctuations.
The steep dip in this gap around $\r=2.5$, together with an increasing coupling 
between the excited states $\chi_n$, given by $\la\chi_n|\partial_\r\chi_m\ra$,
is responsible for the breakdown of the adiabatic approximation dramatically 
illustrated by the sudden increase of $V_{\eff}$ when approaching $\r=2.5$. 
In Fig.~\ref{fig:Ers} we also show $E_2(\r)$ to illustrate that the kink 
in $E_1(\r)$ around $\r=1.7$ is due to an avoided level crossing.

When approaching $\r=0$, the energies of all the excited states grow as 
$\r^{-2}$, due to the non-zero angular momentum, but $E_0(\r)$ goes smoothly 
to zero, as does $\la\partial_\r\chi_0|\partial_\r\chi_0\ra/2$. For 
$\r\rightarrow0$ one easily finds $\chi_0=2\sqrt{\pi}\,\hat\cY_1=2\sqrt{\pi}\,
\hat\cY_0^{001}$, and $\r^{-3}f_0(\r)$ will approach a constant (cmp. the 
behavior of $\phi_p^3(\r)$ introduced in Eq.\Ref{eq:besj}). Near $\r=0$ it 
would be more appropriate to define $\tilde V_\eff(\r)=E_0(\r)+\half A^2(\r)$ 
as the effective potential, shown in Fig.~\ref{fig:Ers} as the dotted curve. 

\subsection{Groundstate energy and wave function}\label{subsec:grst}
The boundary condition to be imposed should be such that $f_0(\r)$ goes to a 
constant  for large $\r$, and we therefore impose $\partial_\r f_0(\r)=0$ at 
the boundary of the fundamental domain, $\r=b\equiv\pi g^{-2/3}(L)$. This is 
equivalent to $\la\chi_0|\partial_\r(\r^{3}\Psi)\ra=0$, where the inner product
is at fixed $\r$. Assuming also that $\la\chi_n|\partial_\r(\r^{3}\Psi)\ra=0$ 
for {\em all} $n$, we can conclude $\partial_\r(\r^{3}\Psi)=0$, and the 
boundary condition translates into a condition on the radial momenta, 
$\partial_\r j_\ell(k b)=0$. The $b$ dependence of these momenta and 
the associated radial matrix elements scale with a simple power of $b$. To 
obtain the full matrix of the Hamiltonian for arbitrary $b$, it suffices 
therefore to calculate momenta and matrix elements at $b=1$. The relevant 
matrix elements, $<p',\ell'|\r^t|p,\ell>$ can be computed either numerically 
or following the algorithm developed earlier for the bosonic ($F=0$) 
case.\cite{Kovb} Needed are the matrix elements with $t=1~(|\ell'-\ell|=1)$
and $t=4~(|\ell'-\ell|=0,2,4)$, as well as $t=2,5$ and $8$ (with appropriate 
selection rules), that occur in determining $H^2$ needed for Temple's 
inequality. As compared to the bosonic case,\cite{Kovb} a change in the 
boundary condition is due to the spherical approximation we have used. In 
reality the boundary is not a sphere but a torus, with an appropriately 
different decomposition of the wave function along the vacuum valley. The 
spherical approximation is justified in the supersymmetric case, since 
$f_0(\r)$ becomes constant well before we reach the boundary.

In the numerical determination of the groundstate energy for the zero-momentum 
Hamiltonian, there are two reasons this boundary cannot be chosen too far from 
the origin. The first reason is, as seen for $E_0(\r)$, that it would require 
too many spherical harmonics to properly localize the wave function in the 
vacuum valley for large $b$. The second reason is that for increasing $b$ 
the energy gap due to the radial excitations goes to zero, another way of 
expressing the fact that for $b\rightarrow\infty$ the spectrum becomes 
continuous down to zero energy.\cite{LNdW} To reach $b=5$ we have used 
420 spherical harmonics, and for each up to 20 radial modes. To keep the 
size of the matrix manageable, the components of the eigenvectors are removed 
when in absolute value below a threshold (typically $\sim10^{-5}$), without 
significantly affecting the accuracy. This process of {\em pruning} is 
performed iteratively, increasing the number of radial modes per angular 
state, for those harmonics that stay above the threshold. Together with 
Temple's inequality this is extremely efficient to optimize the accuracy 
and achieve numerical control. The number of basis vectors needed increases 
with $b$, and ranged up to about 3000 to achieve $|\delta E_0|<0.003$ as 
estimated from Temple's inequality (typically the accuracy of the upper 
bound is much better than this, which we estimate to be $|\delta E_0|<0.0001$).

\begin{figure}[htb]
\vspace{5.6cm}
\includegraphics{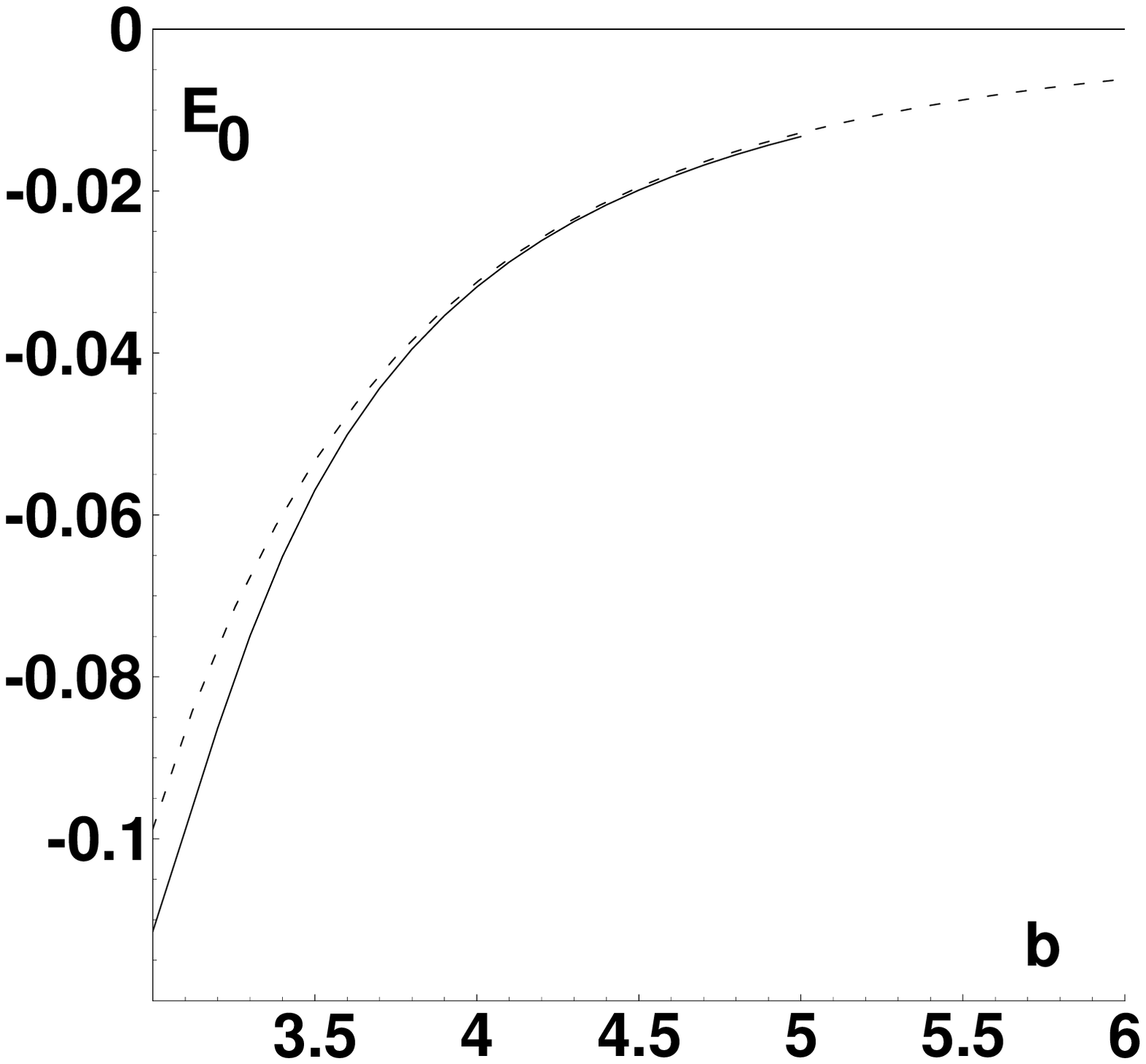}
\includegraphics{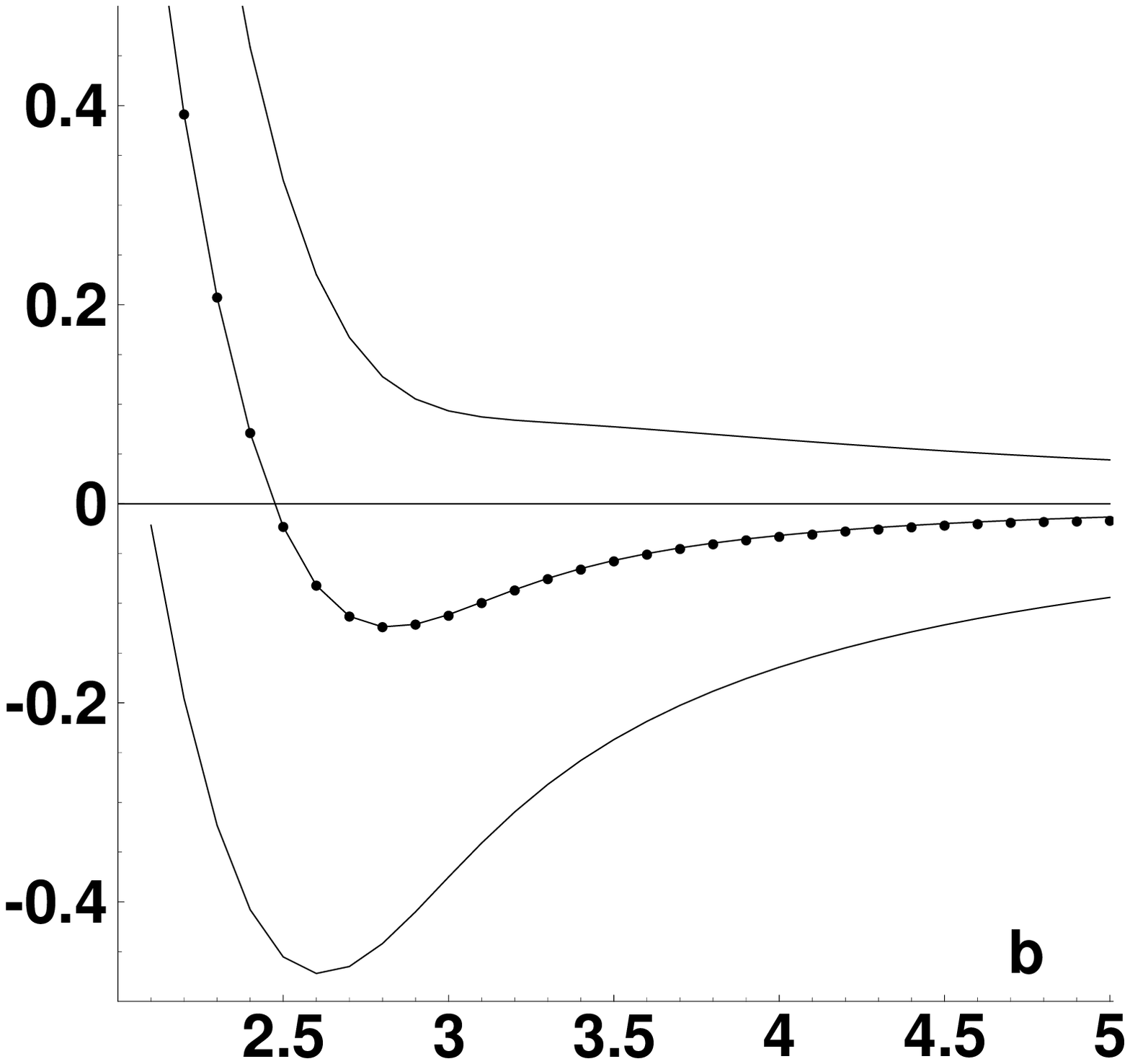}
\caption{The groundstate energy $E_0(b)$, using up to 20 radial modes for each 
of the 420 harmonics. On the left is shown at an enlarged scale the result
obtained with the appropriate boundary condition, $\partial_\r(\r^3\Psi)(b)=0$ 
($\rho=0$). The dashed curve is for $-8/b^4$. The same result is shown on the 
right, together with the lower bound from Temple's inequality (indicated by
the dots), and in comparison to inappropriate choices of boundary conditions
(top curve $\rho=1$, lower curve $\rho=-1$).}\label{fig:Ebs}
\end{figure}

It is crucial to note that for $b$ finite our boundary condition breaks 
supersymmetry. In such a case the groundstate energy need not be positive. 
To illustrate this, and the sensitivity to the boundary conditions, we 
consider in Fig.~\ref{fig:Ebs} the groundstate energy for $\partial_\r(\r^{3+
\rho}\Psi)(b)=0$, with $\rho=-1,0,1$. Indeed, $\rho=0$ makes the groundstate 
energy approach zero most efficiently. It would be tempting to conclude this 
approach is exponential, but our numerical results rather seem to imply 
$E_0(b)\sim -8/b^4$, as indicated by the dashed curve in this figure. The 
boundary condition $\partial_\r f_0(b)=0$ indeed receives perturbative 
corrections, due to the non-vanishing of $\partial_\r\chi_0(b)$.

However, this is an {\em artifact} of the truncation to the zero-momentum modes.
If we take into account that in the full theory $\chi_0$ also involves the 
non-zero momentum modes, the (gauge) symmetry guarantees\Cite{Kovb} that 
{\em at the boundary} of the fundamental domain $\partial_\r\chi_0(b)=0$, 
and this source of the breaking of supersymmetry is absent. The groundstate 
energy will in this case vanish to all orders in perturbation theory. Higher 
order terms in computing the effective Hamiltonian are required to deal with 
this, and was the reason for investigating more closely the determination of 
the effective Hamiltonian. Corrections involving derivatives are manifestations
of non-adiabatic behavior, but they come from the non-zero momentum modes and 
can be treated perturbatively (even when some of these corrections no longer
respect the spherical symmetry). They were not included here.

\begin{figure}[htb]
\vspace{4.7cm}
\includegraphics{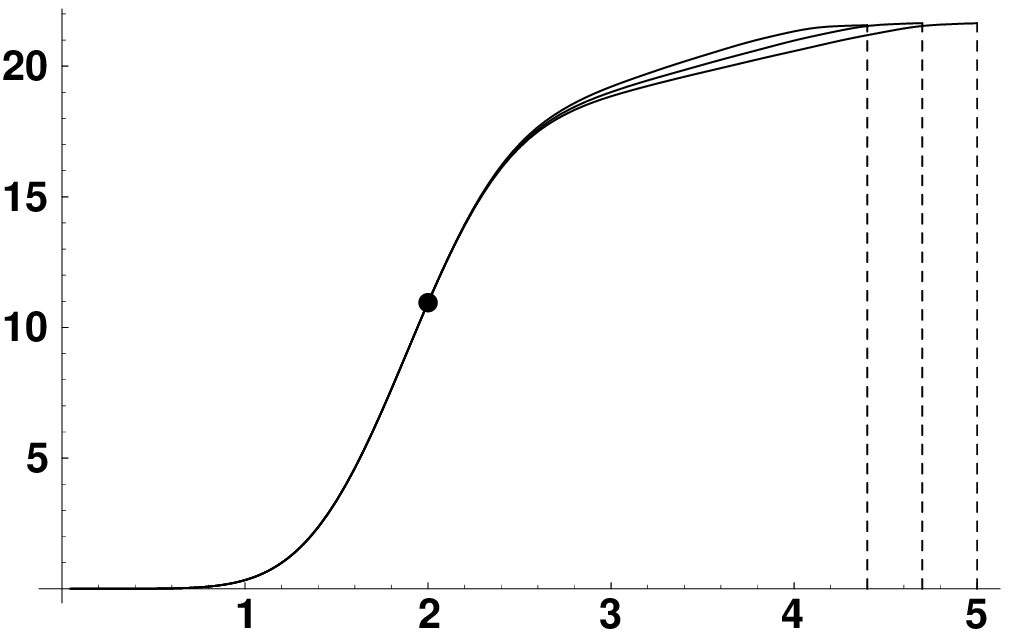}
\includegraphics{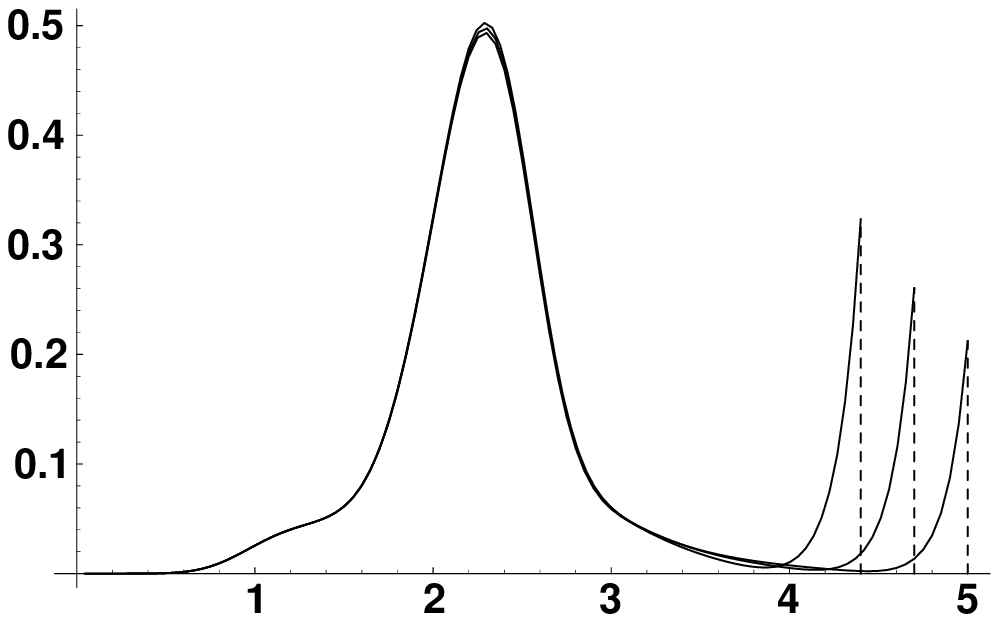}
\vskip-4.2cm \hskip.5cm ${\bf\times 10^{-4}}$ \hskip4.95cm 
${\bf\times 10^{-4}}$\hfill\vskip.4cm\hskip.55cm ${\bf f^2}$
\hskip5.7cm ${\bf f^2\!\!-\!f_0^2}$ \hfill \vskip2.2cm \hskip4cm 
${\bf\rightarrow\hat r}$\hskip4cm ${\bf\rightarrow\hat r}$\vskip0.4cm
\caption{The functions $f^2(\r)$ (left) and $f^2(\r)-f_0^2(\r)$ (right -- 
note the 40-fold increase in scale) extracted from the groundstate wave 
function $\Psi_0$, satisfying the boundary condition $\partial_\r(\r^3\Psi_0)
(b)=0$, for $b=4.4$, 4,7 and 5.0. We normalized with respect to $b=5$, such 
that at $\r=2$ all $f^2$ agree (indicated by the dot).}
\label{fig:fns}
\end{figure}

The effect of truncating the Hamiltonian is also clearly seen from the 
behavior of the wave function near the boundary. In Fig.~\ref{fig:fns} we 
consider the groundstate wave function, satisfying the proper boundary 
condition $\partial_\r(\r^3 \Psi_0)(b)=0$, and plot $f^2(\r)\equiv
\sum_{n=0}^\infty f_n^2(\r)$, as well as $f^2(\r)-f_0^2(\r)=\sum_{n=1}^\infty 
f_n^2(\r)$. With the inner product as defined in Eq.\Ref{eq:inchi}, only 
involving angular integrations, we recall that $f_n(\r)=\r^3\la\Psi_0(\r)|
\chi_n(\r)\ra$ (which can be chosen to be real), $f^2(\r)=\r^6\la\Psi_0(\r)|
\Psi_0(\r)\ra$ and $\la\Psi_0|\Psi_0\ra=\int_0^b 4\pi\r^2f^2(\r)d\r=1$. In 
the adiabatic, or Born-Oppenheimer approximation $f(\r)$ would equal $f_0(\r)$.
A direct measure for the failure of this approximation is given by $f^2(\r)-
f^2_0(\r)$, which as expected deviates from zero when $E_1(\r)-E_0(\r)$ is 
small (cmp. Fig.~\ref{fig:Ers}), but {\em also} when we approach the boundary 
at $\r=b$. In Fig.~\ref{fig:fns} we show the results for $b=4.4$, 4.7 and 5.0, 
to illustrate that this deviation decreases with increasing $b$ (decreasing 
coupling, or volume). The results were normalized such that for all $b$, 
$f^2(\r=2)$ is equal to its value at $b=5$. This shows that the mismatch 
between the boundary condition and the truncation of the effective 
Hamiltonian does not affect the wave function in the neighborhood of the 
orbifold singularities, where the failure of the adiabatic approximation is 
non-perturbative. At the same time we see that beyond this neighborhood of 
the orbifold singularities, $f$ will become constant for large $b$, compatible 
with a vanishing groundstate energy.

\section{Concluding remarks}
Our analysis shows that, although the orbifold singularities are cause for 
concern, in the end they do not upset the result for the Witten index. The 
effective Hamiltonian obtained by reducing to the moduli space of flat 
connections, i.e. the vacuum valley parametrized by the {\em abelian}
zero-momentum modes, requires modification due to a {\em singularity} in the 
non-adiabatic behavior at the orbifold singularities. One is led to consider 
the effective Hamiltonian in the full {\em non-abelian} zero-momentum sector. 
This removes the singularity. {\em Non-singular} corrections due to the 
non-zero momentum modes are still to be included in perturbation theory. 
Supersymmetry is, as usual, expected to keep these perturbative corrections in 
check. In gauge theory the vacuum valley is compact, which can be effectively 
dealt with by imposing boundary conditions in field space. Restricting to this 
fundamental domain is essential, since the non-zero momentum modes will give 
rise to singular non-adiabatic behavior at the other orbifold singularities.
These are gauge copies of $A=0$ and hence outside the fundamental domain. The
boundary conditions can be argued to preserve the supersymmetry if the 
effective Hamiltonian is constructed to all orders in perturbation theory.
Despite all similarities, there is an important distinction with the problem 
of the supermembrane, where the truncated Hamiltonian is assumed to be exact 
(apart from approximating SU($\infty$) by SU(2)) and a boundary condition at a 
{\em finite} distance from the origin will always break the supersymmetry. This 
distinction is a subtle, but important one. 

Of course, a numerical analysis can never be entirely conclusive in deciding a 
theoretical issue that involves the counting of {\em exact} zero-energy states. 
Nevertheless, numerical methods do allow us to quantify any non-perturbative 
contributions, which analytically are out of control due to the orbifold 
singularities. We have shown that indeed these non-perturbative effects do 
not contribute to the vacuum energy, which thus remains zero. In this paper
we have only considered SU(2), to illustrate how to go beyond the adiabatic
approximation. There is no fundamental obstacle to consider other groups, 
including dealing with the newly found disconnected vacuum components, but 
technically this will be much more demanding.

An additional motivation to push ahead with this approach was that our methods 
and results may also be relevant for more general situations in which the 
zero-momentum Hamiltonian (in its truncated form) seems to have a role to 
play. To this purpose we carefully documented our computer code, and make 
it available through the World Wide Web.\cite{Code}

\section*{Acknowledgments}
\addcontentsline{toc}{section}{\numberline{}Acknowledgments}

I thank Daniel Nogradi for a collaboration in the early stages, which resulted
in the Appendix. I also acknowledge fruitful (recent {\em and long past}) 
discussions with Jos\'e Barbon, Jan de Boer, Bernard de Wit, Margarita 
Garc\'{\i}a P\'erez, Marty Halpern, Hiroshi Itoyama, Arjan Keurentjes, Martin 
L\"uscher, Michael Marinov, Hermann Nicolai, Misha Shifman, Andrei Smilga, 
Arkady Vainshtein, John Wheater, Ed Witten and Jacek Wosiek. Finally, I am 
grateful to Maarten Golterman and Steve Sharpe for inviting me to the 
INT-01-3 program on ``Lattice QCD and Hadron Phenomenology", and I thank 
the Institute for Nuclear Theory at the University of Washington in Seattle 
for its hospitality and partial financial support during the crucial phase 
of drafting this paper.

\section*{Appendix}
\addcontentsline{toc}{section}{\numberline{}Appendix}

We can also diagonalize $H_f(\c)$ by solving the system
$\tilde\cS=E_f\cS$ {\em and} $\tilde\cV=E_f\cV$ (see Eq.\Ref{eq:tSV}).
Assuming $E_f\neq0$, we solve $E_f\tilde\cV=E_f^2\cV$ by replacing
$E_f\cS$, as it appears in $E_f\tilde\cV$, by $\tilde\cS$ (which does
{\em not} contain $\cS$). This gives a linear system of equations for 
$\cV$, which is however quadratic in $E_f$,
\be
\cM_{ki}^{ab}\cV_i^b\equiv 2\c_a^k\c^b_i\cV_i^b+\c^a_i\c_k^b\cV_i^b+\c_k^b
\c_i^b\cV_i^a+E_f\veps_{ijk}\veps^{abd}\c^d_j\cV_i^b=E_f^2\cV_k^a.
\ee
For $\c_i^a=\diag(x_1,x_2,x_3)$ (no summations over repeated indices)
\be
\cM_{ki}^{ab}=2x_ix_k\delta_{ka}\delta{ib}+x_ik_k\delta_{ai}\delta_{kb}+x_k^2
\delta_{ki}\delta_{ab}+E_f\veps_{kij}\veps_{abj}x_j,
\ee
which splits in the $3\times3$ block $\cM_0^{ki}=\cM^{ki}_{ki}=2x_i^2
\delta_{ki}+2x_ix_k+E_f(\prod_jx_j)/(x_ix_k)$ and three $2\times2$ blocks 
(forming a triplet under the $x_i$ permutation symmetry) $\cM_j^{ki}=
\cM^{ki}_{ab}=x_ix_k+E_fx_j-E_fx_j\delta_{ik}$, ($a,b$ are fixed by 
requiring $\veps_{kaj}\veps_{ibj}\neq0$, which also constrains 
$i,k\neq j$).  Only $\cM_0$ will give rise to invariant states,
\be
\det(\cM_0-E_f^2)=-(E_f^3-4\c^2E_f-16\det\c)(E_f^3-\c^2E_f+2\det\c)=0,
\ee
agreeing with Eq.\Ref{eq:eigen}. Also note that $\det(\cM_j-E_f^2)=E_f(E_f^3-
\c^2E_f+2\det\c)$ for all $j$, and its resulting energies ($E_f\neq0$) agree 
with the energies of the non-invariant two-gluino states constructed from the 
one-gluino states.\cite{Halp}

\section*{References}
\addcontentsline{toc}{section}{\numberline{}References}

\end{document}